\documentclass[aps,prd,showpacs,preprint,showpacs,showkeys,preprintnumbers,superscriptaddress,amsmath,amssymb,nofootinbib]{revtex4-1}
\usepackage{amsmath}
\usepackage{latexsym}
\usepackage{pstricks,pst-coil}
\usepackage{pst-all}
\def\uma{\rm 1\!\!\hskip 1 pt l}
\usepackage{color}
\usepackage{latexsym}
\usepackage{amsmath}
\usepackage{amssymb}
\usepackage{euscript}
\usepackage{pstricks}
\usepackage{graphics}
\usepackage{graphicx}
\usepackage{picture}

\newcommand{\be}{\begin{equation}}
\newcommand{\ee}{\end{equation}}
\newcommand{\ba}{\begin{eqnarray}}
\newcommand{\ea}{\end{eqnarray}}
\newcommand{\p}{\partial}
\def\ni{\noindent}

\begin{document}

\title{\Large The Strong Interaction Model in DFR noncommutative space-time}

\author{Everton M. C. Abreu}\email{evertonabreu@ufrrj.br}
\affiliation{Grupo de F\' isica Te\'orica e Matem\'atica F\' isica, Departamento de F\'{i}sica, Universidade Federal Rural do Rio de Janeiro, 23890-971, Serop\'edica - RJ, Brasil}
\affiliation{Departamento de F\'{i}sica, Universidade Federal de Juiz de Fora, 36036-330, Juiz de Fora - MG, Brasil}
\author{M. J. Neves}\email{mariojr@ufrrj.br}
\affiliation{Grupo de F\' isica Te\'orica e Matem\'atica F\' isica, Departamento de F\'{i}sica, Universidade Federal Rural do Rio de Janeiro, 23890-971, Serop\'edica - RJ, Brasil}

\date{\today}

\begin{abstract}
\ni The Doplicher-Fredenhagen-Roberts (DFR) framework for noncommutative (NC) space-times
is considered as an alternative approach to describe the physics of quantum gravity. In this formalism,
the NC parameter, {\it i.e.} $\theta^{\mu\nu}$, is promoted to a coordinate of a new extended space-time.
Consequently, we can construct a Field Theory in a space-time with spatial extra-dimensions. This new coordinate
has a canonical momentum associated, where the effects of a new physics point of view can emerge in the fields
propagation along the extra-dimension. In this paper we have introduced the gauge invariance in the DFR NC
space-time by the composite symmetry $U^{\star}(N)\times U^{\star}(N)$.
We have constructed the non-Abelian gauge symmetry based on DFR formalism, and we analyzed the consequences of this symmetry
in the presence of such extra-dimensions. The gauge symmetry in this DFR scenario can reveal new gauge fields
attached to $\theta$-extra-dimension. As an application, we have accomplished the  unification of Strong Interaction with the
electromagnetism and a Higgs model to give masses to the NC bosons. We have estimated their masses by using some
experimental constraints of QCD.

\end{abstract}



\maketitle

\pagestyle{myheadings}
\markright{The Strong Interaction Model in DFR noncommutative space-time}


\section{Introduction}

The inconvenience of having infinities that destroy the final results of several calculations in QFT have motivated theoretical physicists to ask if a continuum space-time would be really necessary. The alternative would be to construct a discrete space-time with a noncommutative (NC) algebra,
where the position coordinates are proportional to operators $\hat{X}^{\mu}\,(\mu=0,1,2,3)$ and they must satisfy the commutation relations
\begin{eqnarray}\label{xmuxnu}
\left[\,\hat{X}^\mu\,,\,\hat{X}^\nu\,\right]\,=\,i\,\ell \, \theta^{\mu\nu} \, \hat{\uma} \,\,,
\end{eqnarray}
where $\ell$ is a length parameter, $\theta^{\mu\nu}$ is an anti-symmetric constant matrix and $\hat{\uma}$, the identity operator. Putting these ideas all together, Snyder \cite{snyder47} published the first work considering the space-time as being a NC one. However, Yang \cite{yang47} demonstrated that Snyder's hopes about the disappearance of the infinities were not achieved by noncommutativity.  This fact has doomed Snyder NC theory to years of ostracism.  After the string theory important result that the algebra obtained from the case of a string theory embedded in a magnetic background is NC, a new wave concerning noncommutativity was rekindle \cite{seibergwitten99}. In current days, the NC approach is also a subject discussed at the quantum gravity level \cite{QG4,QG5}.

One of the paths of introducing noncommutativity is through the Moyal-Weyl (or star) product where the NC parameter,
{\it i.e.} $\theta^{\mu\nu}$, is an anti-symmetric constant. However, at superior orders of calculations, the Moyal-Weyl product turns out to be
highly nonlocal. This fact has lead us to work with low orders in $\theta^{\mu\nu}$. Although it maintains the translational invariance,
the Lorentz symmetry is not preserved \cite{Szabo03}. For example, in the case of the hydrogen atom, it breaks the rotational symmetry of the model, which removes the degeneracy of the energy levels \cite{Chaichian}.

One way to heal this problem was introduced by Doplicher, Fredenhagen and Roberts (DFR) which have promoted the parameter $\theta^{\mu\nu}$ to the role of an ordinary coordinate of the system \cite{DFR1,DFR2}. This so-called extended and new NC space-time has ten dimensions: four
relative to Minkowski space-time ordinary positions and six relative to $\theta$-space. Recently, in \cite{Morita} the authors have conjectured to construct a DFR  space-time extension, introducing the conjugate canonical momentum associated with $\theta^{\mu\nu}$ \cite{Amorim1} (for a review, the reader can see \cite{amo}). This new framework would be characterized by a field theory constructed in a space-time with extra-dimensions $(4+6)$.   Besides, it would
not need necessarily the presence of a length scale $\ell$ localized into the six dimensions
of the $\theta$-space where, from (\ref{xmuxnu}) we can see that $\theta^{\mu\nu}$ would have dimension of length-square (a kind of Planck area),
when we make $\ell=1$. The length scale can be introduced directly in the algebra, and taking the limit with no such scale, the usual algebra of
the commutative space-time is recovered. Besides the Lorentz invariance was also recovered, and obviously we hope that causality aspects in
QFT in this $\left(x+\theta\right)$ space-time must be preserved too, in such way that, the retarded and causal Green's functions of the DFR model are discussed in details in \cite{EMCAbreuMJNeves2012}.

In this approach, as we have said, the parameter $\theta^{\mu\nu}$ is also promoted to a position operator, say $\hat{\theta}^{\mu\nu}$, that participate of the algebra, and it is an observable of the space-time.

In several recent works \cite{Amorim1,Amorim4,Amorim5,Amorim2,EMCAbreuMJNeves2013}, a new version of NC
quantum mechanics (NCQM) were introduced, where not only
the coordinates ${\mathbf x}^\mu$ and their canonical momenta ${\mathbf p}_\mu$ are considered as operators in a Hilbert space ${\cal H}$, but also the objects of noncommutativity $\theta^{\mu\nu}$ and their canonical conjugate momenta $\pi_{\mu\nu}$.
All these operators belong to the same algebra and have the same hierarchical level, introducing  a minimal canonical
extension of DFR formalism. This enlargement of the usual
set of Hilbert space operators allows the theory to be invariant under the rotation group $SO(D)$, as showed
in detail in Ref. \cite{Amorim1,Amorim2}, when the treatment is a non relativistic one.  Rotation invariance in a non relativistic theory is fundamental if one intends to describe any physical system in a consistent way.  It was demonstrated in a precise way that in fact the DFR formalism has a momentum associated with $\theta^{\mu\nu}$.
In the present work we essentially consider the second quantization of the model discussed in Ref \cite{Amorim4}, showing that the extended Poincar\'e symmetry here is generated via generalized Heisenberg relations, giving the same algebra displayed in \cite{Amorim4,Amorim5}. The NC path integral approach and the expectation values of quantum measures are developed in \cite{EMCAbreuMJNeves2013}.

Although we have constructed a NC DFR Klein-Gordon equation \cite{aa} with a source term, an effective action using the Green functions was completely calculated in \cite{EMCAbreuMJNeves2012}. The DFR NC model of scalar field with star self-interaction $\star\phi^4$ was proposed in order to investigate
the divergences at the one loop. It has revealed that at the one loop order, the ultraviolet/infrared divergence mixing phenomena does not emerge in the DFR scalar model $\star\phi^4$, for more details, see \cite{EMCAbreuMJNeves2014}. Posteriorly, the Field Theory formulation has inspired to study the
gauge symmetry in the DFR context when the Dirac/scalar actions are putted under local symmetry \cite{MJNevesEPL2016}. Following these steps,
the composed symmetry $U^{\star}(N) \times U^{\star}(N)$ is the group structure that describes the Yang-Mills scenario in NC DFR space-time.
As consequence, the bosons family is enlarged in relation to usual commutative Yang-Mills model. The immediate application is the version of
Glashow-Salam-Weinberg model by the symmetry $U^{\star}(2) \times U^{\star}(1) \times U^{\star}(2) \times U^{\star}(1)$ whose the extended
Higgs sector gives masses for the new bosons. We estimated these masses using some experimental constraints at the $Z^{0}$ mass, for more details,
see \cite{MJNevesEPL2016}. In this paper, we discuss with more details the extended Yang-Mills symmetry $U^{\star}(N) \times U^{\star}(N)$,
and as application, we propose the unification of Strong Interaction with the electromagnetism DFR through the model $U^{\star}(3) \times U^{\star}(1) \times U^{\star}(3) \times U^{\star}(1)$ . We also introduce a Higgs sector to give masses to some bosons by getting the final symmetry $SU_{c}^{\,\star}(3) \times U^{\star}(1)_{em}$.

The organization of the present paper follows that in section II we have described the DFR formalism.
In section III we have analyzed the charged Klein-Gordon and Dirac equations and their actions. We have introduced new forms and constructions concerning the Gamma matrices in DFR formalism. These new constructions complement those ones given in \cite{Amorim2}. In section IV we study the local gauge
symmetry $U^{\star}(N) \times U^{\star}(N)$. In section V we have dealt with the particular case $U_{em}^{\star}(1)=U^{\star}(1) \times U^{\star}(1)$
that describes the DFR quantum electrodynamics. In section VI we propose the model $U^{\star}(3) \times U^{\star}(1) \times U^{\star}(3) \times U^{\star}(1)$ as the unification of Strong Interaction with electromagnetism. In section VII we construct a Higgs sector to give masses for the NC bosons, and so the enlarged symmetry is reduced to $SU_{c}^{\star}(3) \times U_{em}^{\star}(1)$. In section VIII we estimate the bosons masses using the interactions of the model and some experimental constraints of QCD. The conclusions and perspectives for future works, as always, are depicted in the final section.


\section{The DFR algebra in a nutshell}

In this section, we will construct a NC model based on the one published in \cite{Amorim1,Amorim2,Amorim3,Amorim4,Amorim5}.
Namely, we will revisit the basics of the quantum field theory like those defined DFR space.
The spatial coordinates
of $\theta^{\mu\nu}$ are promoted to quantum observable $\hat{\Theta}^{ij}$ in the commutation relation (\ref{xmuxnu}),
so the spatial non-commutativity is given by
\begin{eqnarray}\label{algebraDFR}
\left[ \, \hat{X}^{i} \, , \, \hat{X}^{j} \, \right] = i \, \hat{\Theta}^{ij} \; ,
\end{eqnarray}
while the time operator does commute with spatial operator usually
\begin{eqnarray}
\left[ \, \hat{X}^{0} \, , \, \hat{X}^{i} \, \right] = 0 \; .
\end{eqnarray}
By convenience, we use the dual operator of $\hat{\Theta}^{ij}$ as our NC
coordinate $\hat{\Theta}^{ij}=\epsilon^{ \, ijk} \, \hat{\Theta}^{k}$.
Moreover there exist the canonical conjugate momenta operator $\hat{K}^{ij}=\epsilon^{\, ijk} \, \hat{K}^{k}$ associated\footnote{The standard notation to represent the momentum is $\pi_{\mu\nu}$ but, for future to-avoid-confusion convenience, we will use from now on, $\hat{K}^{\mu\nu}$.} with the operator $\hat{\Theta}^{ij}$, and they must satisfy the commutation relation
\begin{equation}\label{thetapicomm}
\left[ \, \hat{\Theta}^{i} \, , \, \hat{K}^{j} \, \right] = i \, \delta^{ij} \, \hat{\uma} \; .
\end{equation}
%
In order to obtain consistency we can write that \cite{Amorim1}
\begin{eqnarray}\label{algebraDFRextended}
\left[ \, \hat{X}^{\mu} \, , \, \hat{P}^{\nu} \, \right] = i\,\eta^{\mu\nu} \, \hat{\uma}
\hspace{0.2cm} , \hspace{0.2cm}
\left[ \, \hat{P}^{\mu} \, , \, \hat{P}^{\nu} \, \right] = 0
\hspace{0.2cm} , \hspace{0.2cm}
\left[ \, \hat{\Theta}^{ij} \, , \, \hat{P}^{\rho} \, \right] = 0
\hspace{0.2cm} , \hspace{0.2cm}
\nonumber \\
\left[ \, \hat{P}^{\mu} \, , \, \hat{K}^{ij} \, \right] = 0
\hspace{0.2cm} , \hspace{0.2cm}
\left[\, \hat{X}^{i} \, , \, \hat{K}^{jk}\right]=- \, \frac{i}{4} \, \delta^{ij} \, \hat{P}^{k} + \frac{i}{4} \, \delta^{ik} \, \hat{P}^{j}  \; ,
\end{eqnarray}
and these relations complete the DFR extended algebra\footnote{Here we have adopted that $c=\hbar=\ell=1$, where the $\theta$-coordinate has area dimension.}.
The uncertainty principle from (\ref{xmuxnu}) is modified by
\begin{eqnarray}\label{uncertainxmu}
\Delta \hat{X}^{i} \Delta \hat{X}^{j} \simeq \langle \hat{\Theta}^{ij}\rangle \; ,
\end{eqnarray}
where the expected value of the operator $\hat{\Theta}$ is related to the fluctuation position
of the particles, an it has dimension of length-squared.

The last commutation relation in Eq. (\ref{algebraDFRextended}) suggests that the shifted coordinate
operator \cite{Chaichian,Gamboa,Kokado,Kijanka,Calmet1,Calmet2}
\begin{equation}\label{X}
\hat{\xi}^{i} = \hat{X}^{i}\,+\,{\frac 12} \, \hat{\Theta}^{ij} \, \hat{P}^{j} \,\,,
\end{equation}
commutes with $\hat{K}^{ij}$.
The commutation relation (\ref{algebraDFR}) also commutes with $\hat{\Theta}^{ij}$ and $\hat{\xi}^{i}$,
and satisfies a non trivial commutation relation with $\hat{P}^{\,\mu}$ dependent objects, which could be derived from
\begin{equation}\label{Xpcomm}
\left[ \, \hat{\xi}^{\mu} \, , \, \hat{P}^{\nu} \, \right]=i \, \eta^{\mu\nu} \, \hat{\uma}
\hspace{0.6cm} , \hspace{0.6cm}
\left[ \, \hat{\xi}^{\mu} \, , \, \hat{\xi}^{\nu} \, \right]=0 \,\, ,
\end{equation}
and we can note that the property $\hat{P}_{\mu} \, \hat{\xi}^{\mu}=\hat{P}_{\mu} \, \hat{X}^{\mu}$ is easily verified.
Hence, we can see from these both equations that the shifted coordinated operator (\ref{X}) allows us to recover
the commutativity.
So, differently from $\hat{X}^{i}$, we can say that $\hat{\xi}^{i}$ forms a basis in
Hilbert space.
%
%
%
%
%
%

%
The Lorentz generator group is
\begin{eqnarray}\label{Mmunu}
\hat{M}_{\mu\nu}=\,\hat{\xi}_{\mu} \, \hat{P}_{\nu}\,-\,\hat{\xi}_{\nu} \, \hat{P}_{\mu}
+\hat{\Theta}_{\nu\rho} \, \hat{K}^{\rho}_{\; \;\mu}-\hat{\Theta}_{\mu\rho} \, \hat{K}^{\rho}_{\; \;\nu} \; ,
\end{eqnarray}
and from (\ref{algebraDFRextended}) we can write the generators
for translations as $\hat{P}_{\mu} \longrightarrow \, i \, \partial_{\mu}$.
With these ingredients it is easy to construct the commutation relations
\begin{eqnarray}\label{algebraMP}
\left[ \, \hat{P}_\mu \, , \, \hat{P}_\nu \, \right] &=& 0
\hspace{0.2cm} , \nonumber \\
\left[ \, \hat{M}_{\mu\nu} \, , \, \hat{P}_{\rho} \right] &=& \,i\,\big( \, \eta_{\mu\rho} \, \hat{P}_\nu
-\eta_{\nu\rho} \, \hat{P}_\mu \, \big) \; ,
\hspace{0.1cm} \nonumber \\
\left[ \, \hat{M}_{\mu\nu} \, , \, \hat{M}_{\rho\sigma} \, \right] &=& i\left( \, \eta_{\mu\sigma} \, \hat{M}_{\rho\nu}
-\eta_{\nu\sigma} \, \hat{M}_{\rho\mu}-\eta_{\mu\rho} \, \hat{M}_{\sigma\nu}+\eta_{\nu\rho} \, \hat{M}_{\sigma\mu} \, \right) \; ,
\end{eqnarray}
which closes the proper algebra.  We can say that $\hat{P}_\mu$ and $\hat{M}_{\mu\nu}$
are the DFR algebra generators.

Therefore, we have the structure of a plane space-time $4D$ NC attached to a extra-dimensional $3D$ of spatial coordinates $\{ \theta^{i} \}$
compactified in a sphere $S^{2}$ whose ratio to the square is conjectured by the length scale of the non-commutativity. We construct
the measure of this space-time by means of the line element
\begin{eqnarray}
ds^{2}=\eta_{\mu\nu} \, dx^{\mu} \, dx^{\nu}+\frac{e^{-\frac{\theta_{1}^{\,2}}{4\lambda^{4}}}}{\lambda^{2}} \, \left(d\theta_{1} \right)^{2}
+\frac{e^{-\frac{\theta_{2}^{\,2}}{4\lambda^{4}}}}{\lambda^{2}} \, \left(d\theta_{2} \right)^{2}
+\frac{e^{-\frac{\theta_{3}^{\,2}}{4\lambda^{4}}}}{\lambda^{2}} \, \left(d\theta_{3} \right)^{2} \; ,
\end{eqnarray}
where $\eta^{\mu\nu}=\mbox{diag}(1,-1,-1,-1)$ is the usual Minkowski metric. The sector of extra-dimension is compactified by
gaussian functions that maintain the isotropy of the $\theta$-space. Here we get a matrix $7 \times 7$ as the metric of space-time
$4+3$
\begin{eqnarray}\label{metricg}
g=\mbox{diag}\left(1 \, , \, -1 \, , \, -1 \, , \, -1 \, , \, e^{-\frac{\theta_{1}^{\, 2}}{4\lambda^{4}}} \, , \, e^{-\frac{\theta_{2}^{\, 2}}{4\lambda^{4}}}
\, , \, e^{-\frac{\theta_{3}^{\, 2}}{4\lambda^{4}}} \right) \; .
\end{eqnarray}
It is immediate that the limit $\lambda \rightarrow 0$, the line element becomes the usual Minkowski space-time by the identity
\begin{eqnarray}
\lim_{\lambda \rightarrow 0} \, \frac{e^{-\frac{\theta_{i}^{\,2}}{4 \, \lambda^{4}}}}{\lambda^{2}} = \delta^{3}\left(\theta_{i}^{\, 2}\right) \; ,
\end{eqnarray}
so the $\theta$-part in (\ref{metricg}) is null, {\it i. e.}, $\delta\left(\theta_{i}^{\, 2}\right) \, d\theta_{i}^{\, 2}=0$. Thus, we obtain
\begin{eqnarray}
\lim_{\lambda \rightarrow 0} \, ds^{2}=\eta_{\mu\nu} \, dx^{\mu} \, dx^{\nu} \; .
\end{eqnarray}
An important point in DFR algebra issue is that the Weyl representation of NC operators obeying
the commutation relations keeps the usual form of the Moyal product. In this case, the Weyl map
is represented by
\begin{eqnarray}\label{mapweyl}
\hat{{\cal W}}(f)(\hat{X},\hat{\Theta})=\int\frac{d^{4}p}{(2\pi)^{4}}
\frac{d^{3}{\bf k}}{(2\pi\lambda^{-2})^{3}} \;
\widetilde{f}(p,k) \; e^{i \, p_{\mu} \hat{X}^{\mu} \, + \, i \, {\bf k} \, \cdot \, \hat{\Theta}} \; .
\end{eqnarray}
The Weyl symbol provides a map from the operator algebra to the functions algebra equipped with a star-product,
via the Weyl-Moyal correspondence
\begin{eqnarray}\label{WeylMoyal}
\hat{f}(\hat{X},\hat{\Theta}) \; \hat{g}(\hat{X},\hat{\Theta})
\hspace{0.3cm} \longleftrightarrow \hspace{0.3cm}
f(x,\theta) \star g(x,\theta) \; ,
\end{eqnarray}
where the star-product $\star$ is defined by
\begin{eqnarray}\label{ProductMoyal}
\left. f(x,\theta) \star g(x,\theta) =
e^{\frac{i}{2}\theta^{ij}\partial_{i}\partial^{\prime}_{j}}
f(x,\theta) \, \, g(x^{\prime},\theta) \right|_{x^{\prime}=x} \; ,
\end{eqnarray}
for any arbitrary functions $f$ and $g$ of the coordinates $(x^{\mu},\theta^{ij})$.
Namely, in both sides of Eq. (\ref{ProductMoyal}) we have that $f$ and $g$ are NC objects since they depend on $\theta^{ij}$.

%
%
%
%
%
%
%

Another important ingredient is the so-called function $W$, which is a $\theta$-integration measure that we have introduced by the geometry of the space-time $4+7$ dimensions.
This weight function is introduced in the context of NC field theory to control divergences of the integration
in the $\theta$-space \cite{Carlson,Morita,Conroy2003,Saxell}. Theoretically speaking, it would permit us to work with series expansions
in $\theta$, {\it i.e.}, with truncated power series expansion of functions of $\theta$.
For any large $\theta^{i}$ it falls to zero quickly so that all integrals are well defined,
in its definitions the normalization condition was assumed when integrated in the $\theta$-space.
The function $W$ should be an even function of $\theta$, that is, $W(-\theta)=W(\theta)$ which implies that
an integration in $\theta$-space be isotropic. All the properties involving the
$W$-function can be seen in details in \cite{Carlson,Morita,Conroy2003,Saxell}.
However, we have to say that the role of the $W$-function in NC issues is not altogether clear among the NC community.
By the definition of the Moyal product (\ref{ProductMoyal}) it is
trivial to obtain the property
\begin{eqnarray}\label{Idmoyalproduct2}
\int d^{4}x\,d^{3}\theta \; W(\theta) \; f(x,\theta) \star g(x,\theta)=
\int d^{4}x\,d^{3}\theta \; W(\theta) \; f(x,\theta) \; g(x,\theta) \; .
\end{eqnarray}
%
%
%
%
The physical interpretation of the average of the components of $\theta^{ij}$, {\it i.e.}
$\langle \theta^{2} \rangle$, is the definition of the NC energy scale \cite{Carlson}
\begin{eqnarray}\label{LambdaNC}
\Lambda_{NC}=\left(\frac{12}{\langle \theta^{2} \rangle} \right)^{1/4}=\frac{1}{\lambda} \; ,
\end{eqnarray}
where $\lambda$ is the fundamental length scale
that appears in the Klein-Gordon (KG) equation (\ref{NCKG}) and in the dispersion relation
(\ref{RelDispDFR}) just below. This approach has the advantage of being unnecessary in order to
specify the form of the function $W$, at least for lowest-order processes.
The study of Lorentz-invariant NC QED,
as Bhabha scattering, dilepton and diphoton production
to LEP data led the authors of \cite{Conroy2003,Carone} to determine the bound
\begin{eqnarray}\label{boundLambda}
\Lambda_{NC} > 160 \; GeV \; \; 95 \% \; C.L. \; .
\end{eqnarray}

\section{Field theory in DFR space: Klein-Gordon and Dirac equations}

The first element of the algebra (\ref{algebraMP}) that commutes with all the others generators $\hat{P}^{\mu}$
and $\hat{M}^{\mu\nu}$ is given by
\begin{eqnarray}
\hat{C}_{1}=\hat{P}_{\mu} \, \hat{P}^{\mu}-\frac{\lambda^{2}}{2} \, \hat{K}_{i} \, \hat{K}_{i} \; .
\end{eqnarray}
This is the first Casimir operator of the algebra (\ref{algebraMP}). Using the coordinate representation, the operators
$\hat{P}^{\mu}$ and $\hat{K}_{i}$ can be written in terms of the derivatives
\begin{eqnarray}\label{repcoordinateppi}
\hat{P}_{\mu} \longmapsto i \, \p_{\mu}
\hspace{0.6cm} \mbox{and} \hspace{0.6cm}
\hat{K}_{i} \longmapsto i \, \frac{\p}{\p \theta^{i}}= i \, \p_{\theta i} \; ,
\end{eqnarray}
and consequently, the first Casimir operator in the on-shell condition leads
us to the KG equation in DFR space concerning the scalar field $\phi$
\begin{eqnarray}\label{NCKG}
\left( \, \Box +\lambda^2 \, \Box_\theta+m^2 \, \right)\phi(x,\theta)=0 \; ,
\end{eqnarray}
where we have defined the D'Alembertian $\theta$-operator
\begin{eqnarray}
\Box_{\theta}=\frac{1}{2} \, \p_{ij} \, \p_{ij}=\frac{1}{2} \, \epsilon_{ijk} \, \partial_{\theta k} \, \epsilon_{ijl} \, \partial_{\theta l}
=\partial_{\theta k} \, \partial_{\theta k}=\overrightarrow{\nabla}_{\!\theta}^{\, 2} \; ,
\end{eqnarray}
and $\overrightarrow{\nabla}_{\theta}$ means the $\theta$-gradient operator. The plane wave general solution
for the DFR KG equation is the Fourier integral
\begin{eqnarray}\label{phiXtheta}
\phi( \, x \, , \, \vec{\theta} \, )=\int \frac{d^{4}p}{(2\pi)^{4}} \frac{d^{3}\vec{{\bf k}}}{(2\pi\lambda^{-2})^{3}}
\; \widetilde{\phi}(\, p \, , \, \vec{{\bf k}} \, ) \;  e^{i \, \left( \, p_{\mu} \, x^{\mu}+ \vec{{\bf k}} \, \cdot \, \vec{\theta} \, \right)} \; .
\end{eqnarray}
%
The length $\lambda^{-2}$ is introduced
conveniently in the ${\bf k}$-integration to maintain the field dimension
as being length inverse. Consequently, the $k$-integration keeps dimensionless.
Substituting the wave plane solution (\ref{phiXtheta}), we obtain the invariant mass
\begin{eqnarray}\label{MassInv}
p^{2}-\lambda^{2} \, \vec{{\bf k}}^{\, 2} =m^2 \; ,
\end{eqnarray}
where $\lambda$ is the parameter with length dimension defined before, it
is a Planck-type length.
Thus we obtain the DFR dispersion relation
\begin{eqnarray}\label{RelDispDFR}
\omega(\, \vec{{\bf p}} \, , \, \vec{{\bf k}} \, )=\sqrt{\vec{{\bf p}}^{\,2}
+\lambda^{2}\, \vec{{\bf k}}^{\,2}+m^2} \; .
\end{eqnarray}
%
%
%
It is easy to see that, using the limit $\lambda \rightarrow 0$ in Eqs.
(\ref{NCKG})-(\ref{RelDispDFR}) we can recover the commutative expression
\cite{EMCAbreuMJNeves2012}.
%
%

Since we have constructed the NC KG equation, we will now show its relative action.
We will use the definition of the Moyal-product (\ref{ProductMoyal}) to write the action
for a free complex scalar field in DFR scenario as being
\begin{equation}\label{actionscalarstar}
S_{KG}(\phi^{\ast},\phi)=\int d^{4}x \,d^{3}\theta \, W(\theta) \left( \phantom{\frac{1}{2}} \!\!\!\! \partial_{\mu}\phi^{\ast} \star \partial^{\mu}\phi
+ \lambda^2 \, \partial_{\theta i}\phi^{\ast} \star \partial_{\theta i}\phi
-m^2 \, \phi^{\ast} \star \phi \right) \; ,
\end{equation}
and using the identity (\ref{Idmoyalproduct2}), this free action
can be reduced to the usual one
\begin{eqnarray}\label{actionscalar}
S_{KG}(\phi^{\ast},\phi)=\int d^{4}x \,d^{3}\theta \, W(\theta) \left( \phantom{\frac{1}{2}} \!\!\!\! \left|\partial_{\mu}\phi\right|^{2} +
\lambda^2 \, |\overrightarrow{\nabla}_\theta\phi|^{2}-m^2 \left|\phi\right|^{2} \right) \; .
\end{eqnarray}
\ni where all the products are the usual ones.

It is easy to show that the DFR Dirac equation can be deduced from the square root of the DFR KG equation,
so we can write the field equation \cite{Amorim5}
\begin{eqnarray}\label{EqDiracDFR}
\Big(\,i \, \slash{\!\!\!\partial}
+\,{{i\lambda}\over2}\,\Gamma_{ij} \, \partial_{ij}
-m \, {\uma} \, \Big) \, \Psi(x,\theta)=0 \; ,
\end{eqnarray}
where $\slash{\!\!\!\partial}:=\gamma^{\mu}\partial_{\mu}$, and $\gamma^{\mu}$'s are the ordinary Dirac matrices that satisfy the usual relation
\begin{eqnarray}\label{gammamu}
\left\{ \, \gamma^{\mu} \, , \, \gamma^{\nu} \, \right\}=2\,\eta^{\mu\nu} \; .
\end{eqnarray}
The matrices $\Gamma^{ij}$ are three matrices $4\times4$ which, by construction, they must be anti-symmetric, i.e., $\Gamma^{ij}=-\Gamma^{ij}$.
We can write the matrices $\Gamma^{ij}$ in terms of Dirac matrices-$\gamma$ by means of the commutation relation
\begin{eqnarray}
\Gamma_{ij}:=\frac{i}{2}\left[ \, \gamma_{i} \, , \, \gamma_{j} \, \right]=\sigma_{ij}=\epsilon_{ijk} \, \Sigma_{k} \; ,
\end{eqnarray}
where we can show that the anti-commutation relations are given by
\begin{eqnarray}\label{gammamunu}
\left\{ \, \gamma^{0} \, , \, \Gamma^{jk} \right\}&=&0 \; ,
\nonumber \\
\left\{ \, \gamma^{i} \, , \, \Gamma^{jk} \, \right\}&=&i \, \delta^{i k} \, \gamma^{j}-i \, \delta^{i j} \, \gamma^{k}
+\Gamma^{i j} \, \gamma^{k}-\Gamma^{i k} \, \gamma^{j} \; ,
\nonumber \\
\left \{ \, \Gamma^{ij} \, , \, \Gamma^{kl} \, \right\}&=& \gamma^{k}\, \gamma^{i} \, \delta^{kl}+\gamma^{l} \, \gamma^{j} \, \delta^{ik}
-\gamma^{\lambda} \, \gamma^{i} \, \delta^{jk}
-\gamma^{k}\, \gamma^{j} \,\delta^{il} \; ,
\hspace{1.5cm}
\end{eqnarray}
and the hermiticity property of $\Gamma_{ij}$ is the same as $\gamma^{\mu}$, i.e., $\left(\Gamma_{ij}\right)^{\dagger}=\gamma^{0} \, \Gamma_{ij} \, \gamma^{0}$. This explicit representation
of the matrices $\Gamma_{ij}$ is not a known result in the DFR literature, in which if we use these relations,
the Dirac equation (\ref{EqDiracDFR}) leads us to DFR Klein-Gordon equation.
The components of $\Gamma_{ij}=\epsilon_{ijk} \, \Sigma_{k}$ can be written in terms of the Pauli matrices as
\begin{eqnarray}
%
%
%
\Sigma_{i}=\left(
\begin{array}{cc}
\sigma_{i} & 0 \\
0 & \sigma_{i} \\
\end{array}
\right) \; ,
\end{eqnarray}
which complements the results obtained in \cite{Amorim5}.
%
%
%
%
%
%
The DFR action for the Dirac fermion is
\begin{equation}\label{actionDiracDFRstar}
S_{Dirac}(\bar{\Psi},\Psi)=\int d^{4}x \, d^{3}\theta\, W(\theta) \, \bar{\Psi}(x,\theta) \star
\Big( \, \phantom{\frac{1}{2}} \!\!\!\! \, i \, \slash{\!\!\!\partial}
+\, i \, \lambda \, \overrightarrow{\Sigma} \cdot \overrightarrow{\nabla}_{\theta}
-m \, {\uma} \, \Big) \Psi(x,\theta) \, \,,
\end{equation}
which, using the identity (\ref{ProductMoyal}) can be reduced to
\begin{equation}\label{actionDiracDFRstar2}
S_{Dirac}(\bar{\Psi},\Psi)=\int d^{4}x\,d^{3}\theta\, W(\theta) \, \bar{\Psi}(x,\theta)
\Big(\, \phantom{\frac{1}{2}} \!\!\!\! \, i \, \slash{\!\!\!\partial}
+\, i \, \lambda \, \overrightarrow{\Sigma} \cdot \overrightarrow{\nabla}_{\theta}
-m \, {\uma} \, \Big) \Psi(x,\theta) \, \,,
\end{equation}
which is invariant by symmetry transformations of the Poincar\'e DFR algebra \cite{Amorim5}.
In the next section we will discuss the gauge symmetries of the DFR Dirac Lagrangian.

In the next section we will investigate the coupling of this current with the gauge fields,
which can be an interesting study of the Yang-Mills model in DFR phase-space.
We will also discuss the gauge symmetries of the DFR Dirac Lagrangian.

%

\section{Gauge symmetry and the $U^{\star}(N)\times U^{\star}(N)$ action}

The actions of the complex scalar and Dirac field are invariant under
global transformations of the fields. The invariance of the action under
this global symmetry gives rise to conserved charges, indicated by the conservation law
(\ref{conservlei}). We will now discuss in the invariance of the Dirac action under
local transformations. Let us consider the local gauge transformations for the spinors fields
\begin{eqnarray}\label{Localtransfpsi}
\Psi(x,\theta) \hspace{0.2cm} \longmapsto \hspace{0.2cm}
\Psi^{\prime}(x,\theta)=U(x,\theta)\star\Psi(x,\theta) \; ,
\end{eqnarray}
where $U(x,\theta)$ is an arbitrary matrix $N \times N$ of the coordinates $(x,\theta)$.
It must satisfy the unitarity property
\begin{eqnarray}\label{unitaryU}
U(x,\theta) \star U^{\dagger}(x,\theta)=U^{\dagger}(x,\theta) \star U(x,\theta)=\uma_{N} \; ,
\end{eqnarray}
and we can say that $U$ is {\it star-unitary}, and $\uma_{N}$ is the identity matrix $N \times N$.
%
%
The DFR Dirac Lagrangian is not invariant under the local transformation (\ref{Localtransfpsi}).
To obtain such invariance, we must replace both derivatives $\partial_{\mu}$ and $\lambda\partial_{ij}$
by the following covariant derivatives
\begin{eqnarray}\label{DmuDmunu}
\partial_{\mu} \longmapsto D_{\mu} \, \star &=& \partial_{\mu} - i \, g \, A_{\mu} \star
\nonumber \\
\lambda \, \partial_{\theta i} \longmapsto D_{\theta i} \, \star &=& \lambda \, \partial_{\theta i} - i \, g^{\prime} \, B_{i} \, \star \; ,
\end{eqnarray}
and the NC Dirac fermion Lagrangian is
%
%
%
%
\begin{eqnarray}\label{LDiracDcov}
{\cal L}_{fermion-gauge}=\bar{\Psi}(x,\theta) \star \Big( \, i \, \, \slash{\!\!\!\!D}\star
+\, i \, \overrightarrow{\Sigma} \cdot \overrightarrow{D}_{\theta}\star- m \, {\uma} \, \Big) \Psi(x,\theta)  \; .
\end{eqnarray}
The first one is the usual covariant derivative with a star-product, while $D_{\theta i}\star$ is a
new star-covariant derivative associated with the $\theta$-space
%
%
Consequently, the new field $B_{i} \, , \, i=\{1,2,3\}$
has three independent components
which defines a vector field in the $\left(x+\theta\right)$-space.
This spatial vector field is interpreted like a gauge field associated to compactified $\theta$-space,
that naturally must disappear in the commutative limit, when $\lambda \rightarrow 0$. The quantum
fluctuation of the $B_{i}$ is so attached to $\theta$-space.
The notation $D_{\mu}\star$ indicates a star-product between $A^{\mu}$ and the Dirac spinor, which
does also occur within the covariant derivative $D_{\theta i}\star$ case.
For convenience we have introduced the coupling constants $g$, and $g^{\prime}$
associated with the $\theta$-space. To complete such invariance, we must impose the {\it star-gauge transformations}
\begin{eqnarray}\label{ABgaugetransf}
A_{\mu}(x,\theta) \,\, \longmapsto \,\, A_{\mu}^{\prime}(x,\theta) &=& U(x,\theta)\star A_{\mu}(x,\theta)\star U^{\dagger}(x,\theta)-\frac{i}{g}\left(\partial_{\mu}U\right) \star U^{\dagger}(x,\theta) \; ,
\nonumber \\
B_{i}(x,\theta) \,\, \longmapsto \,\, B_{i}^{\prime}(x,\theta) &=& U(x,\theta) \star B_{i}(x,\theta) \star U^{\dagger}(x,\theta)-\frac{i}{g^{\prime}}\left(\lambda\partial_{\theta i}U\right) \star U^{\dagger}(x,\theta) \; .
\hspace{0.6cm}
\end{eqnarray}

Note that (\ref{unitaryU}) implies that $U^{\dagger}$ is equal to $U^{-1}$ with respect to the star-product
upon the deformed algebra of functions on space-time. In general, it is not true for $\theta \neq 0$,
in which $U^{\dagger} \neq U^{-1}$. An explicit relation between both $U^{\dagger}$ and $U^{-1}$ can be obtained by the series
of the star product, that can be written as
\begin{eqnarray}
U^{\dagger}=U^{-1}+\frac{i}{2} \, \theta^{\mu\nu} \, U^{-1} \, \partial_{\mu}U \, U^{-1} \, \partial_{\nu}U \, U^{-1}
+{\cal O}(\theta^{2}) \; .
\end{eqnarray}
Due to the property $\left(f \star g\right)^{\dagger}=g^{\dagger} \star f^{\dagger}$, the Moyal product $f \star g$
of two unitary matrix fields is always unitary and the group $U^{\star}(N)$ is closed under the star-product.
The special unitary group $SU(N)$ does not give rise to any gauge group in the NC space-time,
because in general $\mbox{det}(f\star g) \neq \mbox{det}(f) \star \mbox{det}(g)$, and consequently,
$\mbox{det}(U \, \star \, U^{\dagger}) \neq \mbox{det}(U) \, \star \, \mbox{det}(U^{\dagger})$, that is, $\det U \neq 1$.

In the opposite case, relative to the
commutative case, $U^{\star}_{N}(1)$ and $SU^{\star}(N)$ are sectors of the decomposition $U^{\star}(N)=U_{N}^{\star}(1) \times SU^{\star}(N)$
do not decouple because the gauge fields of $U_{N}^{\star}(1)$ interacts with the gauge fields of $SU^{\star}(N)$. We represent the $U$-function of $U^{\star}(N)$ as the $\star$-product
\begin{equation}\label{Uexp}
U(x,\theta)=e^{\, i \, \alpha(x,\theta) \, \uma_{N}} \star e^{\, i \, \omega^{a}(x,\theta) \, G^{a}} \; ,
\end{equation}
where $\alpha$, $\omega^{a}$ are arbitrary real functions of $(x,\theta)$ associated with the NC Abelian subgroup
$U_{N}^{\star}(1)$ and with the NC non-Abelian subgroup $SU^{\star}(N)$, respectively, and $G^{a} \, \left( \, a=1 \, , \, 2 \, , \, \cdots \, , \, N^{2}-1 \, \right)$ are the traceless generators of the Lie algebra of the subgroup $SU^{\star}(N)$.

The field $A^{\mu}$ is hermitian gauge field NC
of the star unitary group that we denote by $U^{\star}(N)_{A^{\mu}}$, while that the new field $B_{i}$ is attached
to a second symmetry $U^{\star}(N)_{B^{i}}$, in accord with the transformation (\ref{ABgaugetransf}).
Therefore, the model is constituted by the composite symmetry $U^{\star}(N)_{A^{\mu}} \times U^{\star}(N)_{B^{i}}$
defined in NC space-time scenario.

These fields can be expanded in terms of the Lie algebra
generators of each group $U^{\star}(N)$ correspondent, as for example,
$A_{\mu}=A_{\mu}^{0} \, {\uma_{N}} +A_{\mu}^{a} \, G^{a}$,
and $B_{i}=B_{i}^{0} \, {\uma_{N}} + B_{i}^{a}\, G^{a}$,
with $\mbox{tr}_{N}(G^{a}G^{b})=\delta^{ab}$, $a,b=1,...,N^{2}-1$, by obeying the Lie algebra commutation
relation $\left[G^{a},G^{b}\right]=if^{abc}G^{c}$, and the anti-commutation algebra $\left\{G^{a},G^{b}\right\}=d^{abc}G^{c}$.
The constants $f^{abc}$ and $d^{abc}$ are the structure constants of the Lie algebra.
The fields $A_{\mu}^{0}$ and $B_{i}^{0}$ come from the Abelian part of the correspondent group $U^{\star}(N)$,
while the components $A_{\mu}^{a}$ and $B_{i}^{a}$ are attached
to the non-Abelian part of each $U^{\star}(N)$. More explicitly, the symmetry of model is formed by two abelian
subgroups and two non-abelian subgroups :
\begin{eqnarray*}
U^{\star}(N)_{A^{\mu}} \times U^{\star}(N)_{B^{i}}=U^{\star}(1)_{A^{0\mu}} \times SU^{\star}(N)_{A^{\mu a}} \times U^{\star}(1)_{B^{0i}} \times SU^{\star}(N)_{B^{a i}} \; .
\end{eqnarray*}
Here, the generators $\left\{G^{a}\right\}$ live in the adjoint representation of the groups $U^{\star}(N)$, and $\mbox{tr}_{N}$ denotes the matrix
trace.

Notice that in the fermion sector, the spinor field is the column matrix  whose components are
$\Psi=\left(\psi_{1},\psi_{2}, \cdot \cdot \cdot, \psi_{N}\right)$ that live in the fundamental
representation of the Lie algebra. An important issue is that expressions in NC
gauge theory involve the enveloping algebra of the underlying Lie Group.

The components $A_{\mu}^{\,0}$ and $A_{\mu}^{\,a}$
have the infinitesimal gauge transformation from (\ref{ABgaugetransf})
\begin{eqnarray}
A_{\mu}^{\prime \,\, 0}&=&A_{\mu}^{0}+i\left[ \, \alpha(x,\theta) \, , \, A_{\mu}^{0} \, \right]_{\star}+g^{-1}\partial_{\mu}\alpha(x,\theta) \; ,
\nonumber \\
\nonumber \\
A_{\mu}^{\prime\,\,a}&=&A_{\mu}^{a}-\left[ \, \omega(x,\theta) \, , \, A_{\mu} \, \right]_{\star}^{a}
+i\left[ \, \alpha(x,\theta) \, , \, A_{\mu}^{a} \, \right]_{\star}+i\left[ \, \omega^{a}(x,\theta) \, , \, A_{\mu}^{0} \, \right]_{\star}
+g^{-1}\partial_{\mu}\omega^{a}(x,\theta) \; ,
\end{eqnarray}
and for $B_{i}$, we have that
\begin{eqnarray}
B_{i}^{\prime \, \,0} &=& B_{i}^{0}+i\left[ \, \alpha(x,\theta) \, , \, B_{i}^{0} \, \right]_{\star}+g^{\prime -1}\lambda\partial_{\theta i}\alpha(x,\theta) \; ,
\nonumber \\
\nonumber \\
B_{i}^{\prime \, \, a}&=&B_{i}^{a}-\left[ \, \omega(x,\theta) \, , \, B_{i} \, \right]_{\star}^{a}
+i\left[ \, \alpha(x,\theta) \, , \, B_{i}^{a} \, \right]_{\star}+i\left[ \, \omega^{a}(x,\theta) \, , \, B_{i}^{0} \, \right]_{\star}
+g^{\prime \, -1}\lambda\partial_{\theta i}\omega^{a}(x,\theta) \; .
\nonumber \\
\end{eqnarray}
Using the Moyal product properties, the commutator $\left[ \, \omega \, , \, A_{\mu} \, \right]_{\star}^{a}$ is given by the combination
of the cosine and sine series
\begin{eqnarray}
\label{commutatoromegaA}
\left. \left[ \, \omega(x,\theta) \, , \, A_{\mu}(x,\theta) \, \right]_{\star}^{a}= f^{abc} \, \cos\left(\frac{\theta^{\alpha\beta}}{2}\,\partial_{\alpha}\partial_{\beta}^{\prime} \right)
\omega^{b}(x,\theta) A_{\mu}^{c}(x^{\prime},\theta) \right|_{x^{\prime}=x}
\nonumber \\
\left. + d^{abc} \, \sin\left(\frac{\theta^{\alpha\beta}}{2}\,\partial_{\alpha}\partial_{\beta}^{\prime} \right)
\omega^{b}(x,\theta) A_{\mu}^{c}(x^{\prime},\theta) \right|_{x^{\prime}=x} \; ,
\end{eqnarray}
and the analogous to the case of $\left[ \, \omega \,  , \,  B_{i} \, \right]_{\star}^{a}$.  The simplest commutator
$\left[ \, \alpha \, , \, A_{\mu}^{\; 0} \, \right]_{\star}$ is just the trigonometric sine part of (\ref{commutatoromegaA}),
with $f^{abc}=0$ and $d^{abc}=1$, that goes to zero in the commutative limit.

The $F_{\mu\nu}$ tensor associated with the $A_{\mu}$ field is defined as the {\it star commutation relation}
\begin{eqnarray}\label{DmuDnuFmunu}
\left[ \, D_{\mu} \, , \, D_{\nu} \, \right]_{\star}=-i \, g \, F_{\mu\nu} \; ,
\end{eqnarray}
where
\begin{eqnarray}\label{Fmunu}
F_{\mu\nu}=\partial_{\mu}A_{\nu}-\partial_{\nu}A_{\mu}- i \, g \, \left[ \, A_{\mu} \, , \, A_{\nu} \, \right]_{\star} \; ,
\end{eqnarray}
and it has the gauge transformation
\begin{eqnarray}\label{transfgaugeFmunu}
F_{\mu\nu} \; \longmapsto \; F_{\mu\nu}^{\;\prime} = U(x,\theta) \star F_{\mu\nu} \star U^{\dagger}(x,\theta) \; .
\end{eqnarray}
It is easy to verify that the $F_{\mu\nu}$ tensor is Lie algebra valued of $U^{\star}(N)$ as $F_{\mu\nu}=F_{\mu\nu}^{\,0} \, {\uma_{N}}+F_{\mu\nu}^{a} \, G^{a}$, where the components are given by
%
\begin{eqnarray}\label{ExpF}
F_{\mu\nu}^{0}&=&\partial_{\mu}A_{\nu}^{0}-\partial_{\nu}A_{\mu}^{0}-i \, g \, \left[ \, A_{\mu}^{0} \, , \, A_{\nu}^{0} \, \right]_{\star}
\; ,
\nonumber \\
F_{\mu\nu}^{a}&=&\partial_{\mu}A_{\nu}^{a}-\partial_{\nu}A_{\mu}^{a}+\frac{1}{2} \, g \, f^{abc} \,
\left\{ \, A_{\mu}^{b} \, , \, A_{\nu}^{c} \, \right\}_{\star} -
\nonumber \\
&&-i \, g \, \left[ \, A_{\mu}^{0} \, , \, A_{\nu}^{a} \, \right]_{\star}
-i \, g \, \left[ \, A_{\mu}^{a} \, , \, A_{\nu}^{0} \, \right]_{\star}
-\frac{i}{2} \, g \, d^{abc} \,
\left[ \, A_{\mu}^{b} \, , \, A_{\nu}^{c} \, \right]_{\star} \; .
\end{eqnarray}
These components can be understood as the NC electromagnetic field tensor $F_{\mu\nu}^{0}$, and the Yang-Mills tensor
$F_{\mu\nu}^{a}$ defined in the NC space-time DFR.
%
%
%
Analogously, we obtain the field strength of $B_{i}$
by calculating the star commutation relation
\begin{eqnarray}\label{RelCommutDmunuG}
\left[\, D_{\theta i} \, , \, D_{\theta j} \, \right]_{\star}=-i \, g^{\prime} \, G_{ij} \; ,
\end{eqnarray}
where we obtain
\begin{eqnarray}
G_{ij}=\lambda \, \partial_{\theta i}B_{j}-\lambda \, \partial_{\theta j}B_{i}-i \, g^{\prime}\left[ \, B_{i} \, , \, B_{j} \, \right]_{\star} \; .
\end{eqnarray}
Writing $G_{ij}$ in the basis of $U^{\star}(N)_{B^{i}}$, the components of $G_{ij}$ are
\begin{eqnarray}\label{Gmunu}
G_{ij}^{0}&=&\lambda \, \partial_{\theta i}B_{j}^{0}
-\lambda \, \partial_{\theta j}B_{i}^{0}-i \, g^{\prime} \, \left[ \, B_{i}^{0} \, , \, B_{j}^{0} \, \right]_{\star}
\nonumber \\
G_{ij}^{a}&=&\lambda \, \partial_{\theta i}B_{j}^{a}-\lambda \, \partial_{\theta j}B_{i}^{a}
+ \frac{1}{2} \, g^{\prime} \, f^{abc} \,
\left\{ \, B_{i}^{b} \, , \, B_{j}^{c} \, \right\}_{\star}-
\nonumber \\
&&-i \, g^{\prime}\left[ \, B_{i}^{0} \, , \, B_{j}^{a} \, \right]_{\star}
-i \, g^{\prime}\left[ \, B_{i}^{a} \, , \, B_{j}^{0} \, \right]_{\star}
-\frac{i}{2} \, g^{\prime} \, d^{abc} \,
\left[ \, B_{i}^{b} \, , \, B_{j}^{c} \, \right]_{\star}  .
\end{eqnarray}
%
%
It has the anti-symmetric properties $G_{ij}=-G_{ji}$,
%
%
and the gauge transformation
\begin{eqnarray}\label{transfgaugeG}
G_{ij} \; \longmapsto \;
G^{\;\prime}_{ij}=U(x,\theta) \star G_{ij} \star U^{\dagger}(x,\theta) \; .
\end{eqnarray}
From (\ref{transfgaugeFmunu}) and (\ref{transfgaugeG}), the infinitesimal transformations of the components of $F_{\mu\nu}$ and $G_{\mu\nu\rho\sigma}$ are given by
\begin{eqnarray}\label{TransfinfFG}
F_{\mu\nu}^{0} \; \longmapsto \; F_{\mu\nu}^{0 \, \prime} &=& F_{\mu\nu}^{0}+i \, \left[ \, \alpha \, , \, F_{\mu\nu}^{0} \, \right]_{\star} \; ,
\nonumber \\
F_{\mu\nu}^{a} \; \longmapsto \; F_{\mu\nu}^{a \, \prime} &=& F_{\mu\nu}^{a}-\frac{1}{2}\, f^{abc} \left\{ \, \omega^{b} \, , \, F_{\mu\nu}^{c} \, \right\}_{\star}
+i \, \left[ \, \alpha \, , \, F_{\mu\nu}^{a} \, \right]_{\star}+
\nonumber \\
&&+i \, \left[ \, \omega^{a} \, , \, F_{\mu\nu}^{0} \, \right]_{\star}
+i \, d^{abc}\left[ \, \omega^{b} \, , \, F_{\mu\nu}^{c} \, \right]_{\star} \; ,
\nonumber \\
G_{ij}^{0} \; \longmapsto \; G_{ij}^{0 \, \prime} &=& G_{ij}^{0}+i \, \left[ \, \alpha \, , \, G_{ij}^{0} \, \right]_{\star} \; ,
\nonumber \\
G_{ij}^{a} \; \longmapsto \; G_{ij}^{a \, \prime} &=& G_{ij}^{a}-\frac{1}{2}\, f^{abc} \left\{ \, \omega^{b} \, , \, G_{ij}^{c} \, \right\}_{\star}+i \, \left[ \, \alpha \, , \, G_{ij}^{a} \, \right]_{\star}+
\nonumber \\
&&+i \, \left[ \, \omega^{a} \, , \, G_{ij}^{0} \, \right]_{\star}
+i \, d^{abc}\left[ \, \omega^{b} \, , \, G_{ij}^{c} \, \right]_{\star} \; .
\end{eqnarray}
%
%
%
Therefore we have a Lagrangian for the gauge fields which is invariant under the transformations
(\ref{transfgaugeFmunu}) and (\ref{transfgaugeG})
\begin{equation}\label{LAB}
{\cal L}_{gauge}=
-\frac{1}{4}\,\mbox{tr}_{N}\left(F_{\mu\nu}\star F^{\mu\nu}\right)
-\frac{1}{4}\,\mbox{tr}_{N}\left(G_{ij}\star
G_{ij}\right)
-\frac{1}{2}\,\mbox{tr}_{N}\left(F_{ij} \star G_{ij}
\right)
\; .
\end{equation}
Analogously, this invariance can be applied to the KG Lagrangian of (\ref{actionscalarstar})
by substituting the ordinary derivatives $\partial_{\mu}$ and $\partial_{\theta i}$,
by the covariant derivatives $D_{\mu}$ and $D_{\theta i}$, respectively, we can write
\begin{eqnarray}\label{LescalarDDmunu}
{\cal L}_{KG-gauge}=\left(D_\mu\Phi\right)^{\dagger} \star D^\mu\Phi
+\left(D_{\theta i}\Phi\right)^{\dagger} \star D_{\theta i}\Phi
-m^2 \, \Phi^{\dagger} \star \Phi \; ,
\end{eqnarray}
where the scalar field $\Phi$ represents a multiplet of $N$ complex scalar fields, namely,
$\Phi=\left( \, \phi_{1} \, , \, \phi_{2} \, , \, \cdots \, , \, \phi_{N} \, \right)$.

The DFR version of a NC quantum electrodynamics (QED) is represented when $N=1$.
In this case, the non-abelian subgroup does not contribute for the symmetry,
so we make $G^{a}=0$, $g=e \, Q_{em}$, $g^{\prime}=e^{\prime} \, Q_{em}$.
The gauge field $A^{\mu}$, and the antisymmetrical $B_{ij}$ have as generator the identity matrix,
and the symmetry is composite by
\begin{eqnarray*}
U^{\star}(1)_{em}=U^{\star}(1)_{A^{\mu}} \times U^{\star}(1)_{B^{i}} \; .
\end{eqnarray*}
So the element of each group $U^{\star}(1)$ is given by
\begin{eqnarray}
V(x,\theta)=e^{i \, {\uma} \, \alpha(x,\theta)} \; ,
\end{eqnarray}
and the symmetry is ruled by the transformations
\begin{eqnarray}\label{Localtransfpsi}
\Psi \, \, \longmapsto \, \,
\Psi^{\prime}(x,\theta) &=& e^{i \, \alpha(x,\theta)} \star \Psi(x,\theta) \; ,
\nonumber \\
A_{\mu} \,\, \longmapsto \,\, A_{\mu}^{\prime} &=& V\star A_{\mu}\star V^{\dagger}+\frac{i}{e}\left(\partial_{\mu}V\right) \star V^{\dagger} \; ,
\nonumber \\
B_{i} \,\, \longmapsto \,\, B_{i}^{\prime} &=& V \star B_{i} \star V^{\dagger}+\frac{i}{e^{\prime}}
\left(\lambda\partial_{\theta i}V\right) \star V^{\dagger} \; ,
\end{eqnarray}
where $\Psi$ is a NC spinor of four components, and we have made $Q_{em}=-1$.
%
%
%
The expression of the electromagnetic tensor, and the field strength of $B_{i}$ are given by
%
\begin{eqnarray}\label{ExpFem}
F_{\mu\nu}&=&\partial_{\mu}A_{\nu}-\partial_{\nu}A_{\mu}
+ i \, e \, \left[ \, A_{\mu} \, , \, A_{\nu} \, \right]_{\star} \; \; \; ,
\nonumber \\
G_{ij}&=&\lambda \, \partial_{\theta i}B_{j}
-\lambda \, \partial_{\theta j}B_{i}
+ i \, e^{\prime} \, \left[ \, B_{i} \, , \, B_{j} \, \right]_{\star} \;  ,
\end{eqnarray}
%
%
%
%
where we have achieved an invariance property for the DFR Dirac action under the local transformations
(\ref{Localtransfpsi}). To guarantee this invariance we must introduce an anti-symmetric
field $B_{i}$ in (\ref{DmuDmunu}), beyond the vector field $A^{\mu}$.
The field $A^{\mu}$ has the first transformation of (\ref{ABgaugetransf}), while $B_{i}$ has the second transformation of (\ref{ABgaugetransf}).
This new anti-symmetric field is associated with the $\theta$-space and it must to be attached to those extra-dimensions.
The NC gauge theory obtained here is reduced to the standard case of $SU(N)$ Yang-Mills,
and the usual $U(1)$ QED, in the commutative limit $\theta=0$, and taking $\lambda=0$ in the Lagrangian (\ref{LDiracDcov}),
(\ref{LAB}) and (\ref{LescalarDDmunu}).

In the next section we will discuss the field equations and the currents of the star-symmetry $U^{\star}(1)_{A^{\mu}}\times U^{\star}(1)_{B_{i}}$
in DFR space.
%
%

\pagebreak

\section{The DFR Electromagnetism}

The electromagnetic
equations in DFR space-time will be computed in this section.
To accomplish the task, we have the gauge sector which obeys the star gauge symmetry $U^{\star}(1)_{A^{\mu}}\times U^{\star}(1)_{B_{i}}$
discussed in the previous section
\begin{eqnarray}\label{LSpinor+gauge}
{\cal L}_{em^{\star}}&=&-\frac{1}{4}\,F_{\mu\nu} \, \star \, F^{\mu\nu}
-\frac{1}{4} \, G_{ij} \, \star \, G_{ij}
-\frac{1}{2} \, F_{ij} \, \star \, G_{ij} \; ,
\end{eqnarray}
where the field strength have defined in (\ref{ExpFem}). In the sector of the gauge fields we have naturally the
Maxwell Lagrangian in the context of DFR NC. We can define the components of field strength tensors as
\begin{eqnarray}
F^{\mu\nu}&=&\left(\, {\bf E}^{i} \, , \, \epsilon^{ijk} \, {\bf B}^{k} \, \right)
\nonumber \\
G_{ij}&=&\epsilon_{ijk} \, H_{k} \; ,
\end{eqnarray}
where $H_{i}$ is the dual field of $G_{ij}$, and the Lagrangian more explicitly in terms of these components is
\begin{eqnarray}\label{LEBH}
{\cal L}_{em^{\star}}&=&\frac{1}{2} \left( {\bf E} \star {\bf E}-{\bf B} \star {\bf B}\right)
-\frac{1}{2} \, {\bf H} \star {\bf H}
- {\bf B} \star {\bf H} \; .
\end{eqnarray}

Using the principle of the minimal action associated with respect to $A^{\mu}$, the Lagrangian gives us the NC Maxwell's equations
in the presence of a source $J^{\mu}=\left(\rho,{\bf J}\right)$
\begin{eqnarray}\label{EqfieldA0}
\nabla_{\mu} \, \star \, F^{\,\mu j}
+\epsilon^{\, ijk} \, \, \nabla_{i} \, \star \, H^{\, k} &=& e \, J^{\, j}
\nonumber \\
\nabla_{i} \, \star \, F^{\,i0} &=& e \, \rho \; ,
\end{eqnarray}
where the covariant derivative $\nabla_{\mu}$ acting on strength field tensors is defined by
\begin{eqnarray}\label{nabla0}
\nabla_{\mu} \star F^{\, \mu\nu}:=\partial_{\mu}F^{\, \mu\nu}-e \, \left[ \, A_{\, \mu} \, , \, F^{\, \mu\nu} \, \right]_{\star} \; .
\end{eqnarray}
The tensor $F^{\mu\nu}$ must obey the Bianchi identity
\begin{eqnarray}\label{IDBianchi0}
\nabla_{\mu} \star F_{\nu\rho}+\nabla_{\nu} \star F_{\rho\mu}+\nabla_{\rho} \star F_{\mu\nu}=0 \; ,
\end{eqnarray}
which completes the equations for the NC electromagnetism. The field equations for the $B_{ij}$ tensor fields
are auxiliary equations that emerge exclusively from the NC $\theta$ extra-dimensions
\begin{eqnarray}\label{EqfieldB0}
\epsilon^{ijk} \, \nabla_{\theta}^{\,\, j} \, \star \, H^{k}
+\epsilon^{ijk} \, \nabla_{\theta}^{\,\, j} \, \star \, B^{k}
=-\, e^{\prime} \, \tilde{J}^{\, i} \; ,
\end{eqnarray}
where the anti-symmetric covariant derivative $\nabla_{ij}$ is defined by
\begin{eqnarray}\label{DmunuG}
\epsilon^{ijk} \, \nabla_{\theta}^{\,\, j} \, \star \, H^{k}:=\lambda \, \epsilon^{ijk} \, \partial_{\theta}^{\, j} H^{\, k}
-e^{\prime} \, \, \epsilon^{\, ijk} \,  \left[\, \tilde{B}^{\, j} \, , \, H^{\, k} \,\right]_{\star} \; .
\end{eqnarray}
The current $J^{\,\mu}$ is the classical source of the NC Electric and Magnetic fields, while the
the dual current $\tilde{J}^{k}$ is interpreted as the source of the background field $H^{k}$
due to the extra-dimension of the NC space-time. When the NC parameter goes to zero, the new current $\tilde{J}^{\, k}$
is automatically zero. From the Eqs. (\ref{EqfieldA0}) and (\ref{EqfieldB0}),
we can obtain the conservation law
\begin{eqnarray}\label{EqContinuidade0}
\nabla_{\mu} \star J^{\, \mu} + \nabla_{\theta}^{\,\, k}\star \tilde{J}^{\, k}=0 \; ,
\end{eqnarray}
which expresses the electric charge covariant conservation. The equation (\ref{EqfieldA0})
sets the Ampère Law of the NC magnetic field $B^{k}$, and we have obtained a second Ampère Law
for the background field $H^{k}$. Therefore it can be interpreted as a like background magnetic field
emerged of the $\theta$-space. The gauge symmetry permits us to fix a gauge of Coulomb for the $B_{i}$ field
as
\begin{eqnarray}
\overrightarrow{\nabla}_{\theta} \cdot \overrightarrow{{\bf B}}=0 \; .
\end{eqnarray}

Backing to the Lagrangian (\ref{LSpinor+gauge}) we analyze the interactions that arise due to the NC space.
%
%
%
%
Thus the interactions between NC photons and the field $B^{i}$ have vertex of three and four lines.
The vertex of three lines are
%
%
%
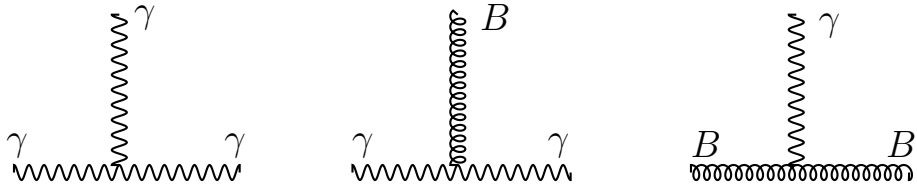
\begin{figure}[!h]
\begin{center}
\newpsobject{showgrid}{psgrid}{subgriddiv=1,griddots=10,gridlabels=6pt}
\begin{pspicture}(0,1.6)(10,3.2)
\psset{arrowsize=0.2 2}
\psset{unit=1}
%
%
%
\pscoil[coilarm=0,coilaspect=0,coilwidth=0.2,coilheight=1.0,linecolor=black](-1.5,1)(1.5,1)
\pscoil[coilarm=0,coilaspect=0,coilwidth=0.2,coilheight=1.0,linecolor=black](-0.1,1.1)(-0.1,3.1)
\put(0.1,3){\large$\gamma$}
\put(-1.6,1.3){\large$\gamma$}
\put(1.3,1.3){\large$\gamma$}
%
%
\pscoil[coilarm=0,coilaspect=0,coilwidth=0.2,coilheight=1.0,linecolor=black](3,1)(5.9,1)
\pscoil[coilarm=0,coilwidth=0.2,coilheight=1.0,linecolor=black](4.4,1.1)(4.4,3.1)
\put(4.7,2.9){\large$B$}
\put(3,1.3){\large$\gamma$}
\put(5.6,1.3){\large$\gamma$}
%
%
\pscoil[coilarm=0,coilwidth=0.2,coilheight=1.0,linecolor=black](7.5,1)(10.4,1)
\pscoil[coilarm=0,coilaspect=0,coilwidth=0.2,coilheight=1.0,linecolor=black](8.9,1.1)(8.9,3.1)
\put(9.2,2.9){\large$\gamma$}
\put(7.5,1.2){\large$B$}
\put(10.1,1.2){\large$B$}
%
%
\end{pspicture}
%
%
%
\end{center}
\end{figure}

\noindent
in which the correspondent Lagrangian is
\begin{eqnarray}\label{Lintvertex3}
{\cal L}_{int}^{\,\,(3)}=- e \, \partial_{\mu}A_{\nu} \, \left[ \, A^{\mu} \, , \, A^{\nu} \, \right]_{\star}
- \lambda \, e \, \partial_{\theta i}B_{j} \, \left[ \, A_{i} \, , \, A_{j} \, \right]_{\star}
-e^{\prime} \, \partial_{i}A_{j} \, \left[ \, B_{i} \, , \, B_{j} \, \right]_{\star} \; .
\end{eqnarray}
For the vertex of four lines we have the diagrams
%
%
\begin{figure}[!h]
\begin{center}
\newpsobject{showgrid}{psgrid}{subgriddiv=1,griddots=10,gridlabels=6pt}
\begin{pspicture}(-2,1)(11,3.5)
\psset{arrowsize=0.2 2}
\psset{unit=1}
%
%
%
\pscoil[coilarm=0,coilaspect=0,coilwidth=0.2,coilheight=1.0,linecolor=black](-1.5,1)(1.5,1)
\pscoil[coilarm=0,coilaspect=0,coilwidth=0.15,coilheight=1.0,linecolor=black](-0.1,1.1)(-1,3)
\pscoil[coilarm=0,coilaspect=0,coilwidth=0.15,coilheight=1.0,linecolor=black](-0.1,1.1)(1,3)
\put(-1.4,2.9){\large$\gamma$}
\put(1.15,2.9){\large$\gamma$}
\put(-1.6,1.3){\large$\gamma$}
\put(1.3,1.3){\large$\gamma$}
%
%
\pscoil[coilarm=0,coilwidth=0.2,coilheight=1.0,linecolor=black](3,1)(5.9,1)
\pscoil[coilarm=0,coilwidth=0.2,coilheight=1.0,linecolor=black](4.4,1.1)(3.4,3)
\pscoil[coilarm=0,coilwidth=0.2,coilheight=1.0,linecolor=black](4.4,1.1)(5.4,3)
\put(2.85,2.9){\large$B$}
\put(5.5,2.9){\large$B$}
\put(3,1.2){\large$B$}
\put(5.6,1.2){\large$B$}
%
%
\pscoil[coilarm=0,coilwidth=0.2,coilheight=1.0,linecolor=black](7.5,1)(10.4,1)
\pscoil[coilarm=0,coilaspect=0,coilwidth=0.15,coilheight=1.0,linecolor=black](8.9,1.1)(8,3)
\pscoil[coilarm=0,coilaspect=0,coilwidth=0.15,coilheight=1.0,linecolor=black](8.9,1.1)(10,3)
\put(7.6,2.9){\large$\gamma$}
\put(10.15,2.9){\large$\gamma$}
\put(7.5,1.2){\large$B$}
\put(10.1,1.2){\large$B$}
%
%
\end{pspicture}
%
%
%
\end{center}
\end{figure}
\pagebreak
\noindent
and the Lagrangian is
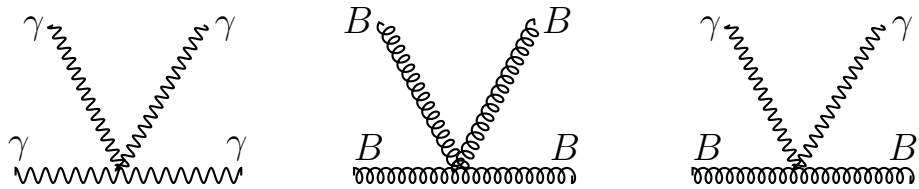
\begin{equation}\label{Lint4AABB}
{\cal L}_{int}^{\,\,(4)}=\frac{e^{2}}{4} \, \left[ \, A_{\mu} \, , \, A_{\nu} \, \right]_{\star}^{\; 2}
+\frac{\left(e'\right)^{2}}{4} \, \left[ \, B_{i} \, , \, B_{j} \, \right]_{\star}^{\; 2}
+\frac{e \, e^{\prime}}{2} \, \left[ \, A_{i} \, , \, A_{j} \, \right]_{\star}
\, \left[ \, B_{i} \, , \, B_{j} \, \right]_{\star} \; .
\end{equation}
\section{The model $U^{\star}(3) \times U^{\star}(1) \times U^{\star}(3) \times U^{\star}(1)$.}
As an application of model $U^{\star}(N) \times U^{\star}(N)$, we investigate the case $N=3$, in which we have the symmetry
$U^{\star}(3) \times U^{\star}(1) \times U^{\star}(3) \times U^{\star}(1)$ as the candidate to unify the strong interaction
to electromagnetism in the DFR NC scenario.
Explicitly, the complete symmetry is represented by the composition
\begin{eqnarray*}
U^{\star}(3)_{A^{\mu}} \times U^{\star}(1)_{X^{\mu}} & \times & U^{\star}(3)_{B^{i}} \times U^{\star}(1)_{Y^{i}}=
\nonumber \\
&&\hspace{-1.0cm}
=U^{\star}(1)_{A^{0\mu}} \times SU^{\star}(3)_{A^{\mu a}} \times  U^{\star}(1)_{X^{\mu}} \times U^{\star}(1)_{B^{0i}} \times SU^{\star}(3)_{B^{a i}} \times U^{\star}(1)_{Y^{i}} \; .
\end{eqnarray*}
The sector of quarks and leptons of this model is composite by
\begin{eqnarray}\label{Lquarks}
{\cal L}_{quarks/leptons}=\bar{q}_{u}\left( \!\!\!\! \phantom{\frac{1}{2}} i \, \, \slash{\!\!\!\!D}\star + \, i \, \overrightarrow{\Sigma} \cdot \overrightarrow{D}_{\theta}\star - m_{u} \, {\uma} \right) q_{u}
\nonumber \\
+ \, \bar{q}_{d}\left( \!\!\!\! \phantom{\frac{1}{2}} i \, \, \slash{\!\!\!\!D}\star + \, i \, \overrightarrow{\Sigma} \cdot \overrightarrow{D}_{\theta}\star - m_{d} \, {\uma} \right) q_{d}
\nonumber \\
+ \, \bar{\ell}\left( \!\!\!\! \phantom{\frac{1}{2}} i \, \, \slash{\!\!\!\!D}\star + \, i \, \overrightarrow{\Sigma} \cdot \overrightarrow{D}_{\theta}\star - m_{\ell} \, {\uma} \right) \ell \; ,
\end{eqnarray}
where the components of covariant derivatives that act on quarks and leptons spinors are defined by
\begin{eqnarray}\label{Dcovqell}
D_{\mu}\star q &:=& \partial_{\mu} \, q+i \, g_{1} \, A_{\mu} \, \star q
- i \, g_{1}^{\prime} \, X \, q \star  X_{\mu}
\nonumber \\
D_{\theta i}\star q &:=&  \lambda \, \partial_{\theta i} \, q + i \, g_{2} \, B_{i} \star q - i \, g_{2}^{\prime} \, Y \, q \star Y_{i}
\nonumber \\
D_{\mu}\star \ell &:=& \partial_{\mu} \, \ell - i \, g_{1}^{\prime} \, X \, \ell \star  X_{\mu}
\nonumber \\
D_{\theta i}\star \ell &:=& \lambda \, \partial_{\theta i} \, \ell - i \, g_{2}^{\prime} \, Y \, \ell \star Y_{i} \; ,
\end{eqnarray}
in which $A^{\mu}$ and $B^{i}$ are Lie group valued of $U^{\star}(3)$ as
\begin{eqnarray}
A^{\mu}&=&{\uma} \,  A^{0\mu} + \, \frac{\lambda^{a}}{2} \, A^{\mu \,a}
\nonumber \\
B^{i}&=&{\uma} \, B^{0i} + \frac{\lambda^{a}}{2} \, B^{i \, a} \; .
\end{eqnarray}
The Gellman matrices $\{\frac{\lambda^{a}}{2}\}$ satisfy the Lie algebra of subgroup $SU^{\star}(3)$
\begin{eqnarray}
\left[ \, \frac{\lambda^{a}}{2} \, , \, \frac{\lambda^{b}}{2} \, \right]=i \, f^{abc} \, \frac{\lambda^{c}}{2} \; ,
\end{eqnarray}
where $\mbox{tr}\left( \lambda^{a} \, \lambda^{b} \right)=2 \, \delta^{ab}$.

The quarks $\left\{ \, q_{u} \, , \, q_{d} \, \right\}$ are triplets that transform in the fundamental representation of $U^{\star}(3)$,
and in the anti-fundamental representation of $U^{\star}(1)$ in accord with
\begin{eqnarray}
q_{u}&=&
\left(
\begin{array}{ccc}
q_{u1} & q_{u2} & q_{u3} \\
\end{array}
\right)^{t}
\; \; \longmapsto \; \;
q_{u}^{\prime} = U \star q_{u} \star V^{-1} \; ,
\nonumber \\
q_{d}&=&
\left(
\begin{array}{ccc}
q_{d1} & q_{d2} & q_{d3} \\
\end{array}
\right)^{t}
\; \; \longmapsto \; \;
q_{d}^{\prime} = U \star q_{d} \star V^{-1} \; ,
\end{eqnarray}
where $U$ is the element of $U^{\star}(3)$, and $V$ is the element of $U^{\star}(1)$.
Under the symmetry $U^{\star}(3)$, the gauge fields have the transformations like (\ref{ABgaugetransf}).
The transformation of leptons $\ell=\{ \, e \, , \, \mu \, , \, \tau \, \}$ is in the anti-fundamental representation with respect to Abelian groups $U^{\star}(1)$
\begin{eqnarray}
\ell
\; \; \longmapsto \; \;
\ell^{\, \prime}= \ell \star V^{-1} \; ,
\end{eqnarray}
and the gauge transformations of vector fields are
\begin{eqnarray}
A_{\mu} \,\, \longmapsto \,\, A_{\mu}^{\prime}&=&U\star A_{\mu}\star U^{-1}+\frac{i}{g_{1}}\left(\partial_{\mu}U\right) \star U^{-1} \; ,
\nonumber \\
B_{i} \,\, \longmapsto \,\, B_{i}^{\prime}&=&U \star B_{i} \star U^{-1}+\frac{i}{g_{2}}\left(\lambda\partial_{\theta i}U\right) \star U^{-1} \; ,
\nonumber \\
X_{\mu} \,\, \longmapsto \,\, X_{\mu}^{\prime}&=& V\star X_{\mu}\star V^{-1}+\frac{i}{g_{1}^{\prime} \, X}\left(\partial_{\mu}V\right) \star V^{-1} \; ,
\nonumber \\
Y_{i} \,\, \longmapsto \,\, Y_{i}^{\prime}&=&V \star Y_{i} \star V^{-1}+\frac{i}{g_{2}^{\prime} \, Y}\left(\lambda \, \partial_{\theta i}V\right) \star V^{-1}
\; ,
\end{eqnarray}
Therefore, the transformations of covariant derivatives under $U^{\star}(3)$ and $U^{\star}(1)$ are
\begin{eqnarray}
D_{\mu}\star q \; \; \longmapsto \; \; \left(D_{\mu}\star q\right)' &=& U \star \left(D_{\mu}\star q\right) \star V^{-1}
\nonumber \\
D_{\theta i}\star q \; \; \longmapsto \; \; \left(D_{\theta i}\star q\right)' &=& U \star \left(D_{\theta i}\star q\right) \star V^{-1}
\nonumber \\
D_{\mu}\star \ell \; \; \longmapsto \; \; \left(D_{\mu}\star \ell\right)' &=& \left(D_{\mu}\star \ell\right) \star V^{-1}
\nonumber \\
D_{\theta i}\star \ell \; \; \longmapsto \; \; \left(D_{\theta i}\star \ell\right)' &=& \left(D_{\theta i}\star \ell\right) \star V^{-1}
\; .
\end{eqnarray}
With all these transformations we get the symmetry of the action associated to the Lagrangian (\ref{Lquarks}).

The sector of gauge fields is formed by eight vector fields $A^{\mu a}\, (a=1,2,\cdots,8)$, one abelian NC $A^{\mu 0}$, eight
antisymmetrical fields $B^{i \, a}$, and one antisymmetrical abelian field $B^{i \, 0}$, both due to extra-dimension of the NC space.
The field strength tensors are defined by the calculation of the commutators involving the covariant derivatives of (\ref{Dcovqell})
\begin{eqnarray}
\left[ \, D_{\mu} \, , \, D_{\mu} \, \right]=i \, g_{1} \, F_{\mu\nu}-i \, g_{1}^{\, \prime} \, X_{\mu\nu} \, {\uma}
\end{eqnarray}
and
\begin{eqnarray}
\left[ \, D_{\theta i} \, , \, D_{\theta j} \, \right]=i \, g_{2} \, B_{ij}-i \, g_{2}^{\, \prime} \, Y_{ij} \, {\uma} \; ,
\end{eqnarray}
where it components are given by
\begin{eqnarray}
F_{\mu\nu}^{\,0}&=&\partial_{\mu}A_{\nu}^{\, 0}-\partial_{\nu}A_{\mu}^{\, 0}+i \, g_{1} \, \left[ \, A_{\mu}^{\, 0} \, , \, A_{\nu}^{\, 0}  \, \right]_{\star}
\nonumber \\
&&
- \, i \, g_{1}^{\, \prime} \, X \, \left[ \, A_{\mu}^{\, 0} \, , \, X_{\nu} \, \right]_{\star}
- \, i \, g_{1}^{\, \prime} \, X \, \left[ \, X_{\mu} \, , \, A_{\nu}^{\, 0} \, \right]_{\star}
+ \, \frac{i \, g_{1}}{4} \, \left[ \, A_{\mu}^{\, a} \, , \, A_{\nu}^{\, a}  \, \right]_{\star}
\nonumber \\
F_{\mu\nu}^{\,a}&=&\partial_{\mu}A_{\nu}^{\, a}-\partial_{\nu}A_{\mu}^{\, a}+\frac{1}{2} \, g_{1} \, f^{abc} \, \left\{ \, A_{\mu}^{\, b} \, , \, A_{\nu}^{\, c}  \, \right\}_{\star}
+ \, i \, g_{1} \, \left[ \, A_{\mu}^{\, a} \, , \, A_{\nu}^{\, 0}  \, \right]_{\star}
\nonumber \\
&&
+ \, i \, g_{1} \, \left[ \, A_{\mu}^{\, 0} \, , \, A_{\nu}^{\, a}  \, \right]_{\star}
- \, i \, g_{1}^{\, \prime} \, X \, \left[ \, A_{\mu}^{\, a} \, , \, X_{\nu}  \, \right]_{\star}
- \, i \, g_{1}^{ \, \prime} \, X \, \left[ \, X_{\mu} \, , \, A_{\nu}^{\, a} \, \right]_{\star}
\nonumber \\
X_{\mu\nu}&=&\partial_{\mu}X_{\nu}-\partial_{\nu}X_{\mu}-i \, g_{1}^{\prime} \, X \, \left[X_{\mu},X_{\nu}\right]_{\star}
\end{eqnarray}
and
\begin{eqnarray}
G_{ij}^{\,0} &=& \lambda \, \partial_{\theta i}B_{j}^{\, 0}-\lambda \, \partial_{\theta j}B_{i}^{\, 0}+i \, g_{2} \, \left[ \, B_{i}^{\, 0} \, , \, B_{j}^{\, 0} \, \right]_{\star}
\nonumber \\
&&
- \, i \, g_{2}^{\, \prime} \, Y \, \left[ \, B_{i}^{\, 0} \, , \, Y_{j} \, \right]_{\star}
- \, i \, g_{2}^{\, \prime} \, Y \, \left[ \, Y_{i} \, , \, B_{j}^{\, 0} \, \right]_{\star}
+ \, \frac{i \, g_{2}}{4} \, \left[ \, B_{i}^{\, a} \, , \, B_{j}^{\, a}  \, \right]_{\star}
\nonumber \\
G_{ij}^{\,a}&=&\lambda \, \partial_{\theta i}B_{i}^{\, a}-\lambda \, \partial_{\theta j}B_{j}^{\, a}+\frac{1}{2} \, g_{2} \, f^{abc} \, \left\{ \, B_{i}^{\, b} \, , \, B_{j}^{\, c}  \, \right\}_{\star}
\nonumber \\
&&
+ \, i \, g_{2} \, \left[ \, B_{i}^{\, 0} \, , \, B_{j}^{\, a}  \, \right]_{\star}
- \, i \, g_{2}^{\, \prime} \, Y \, \left[ \, B_{i}^{\, a} \, , \, Y_{j}  \, \right]_{\star}
- \, i \, g_{2}^{ \, \prime} \, Y \, \left[ \, Y_{i} \, , \, B_{j}^{\, a} \, \right]_{\star}
\nonumber \\
Y_{ij}&=&\lambda \, \partial_{\theta i}Y_{j}- \lambda \, \partial_{\theta j}Y_{i}- i \, g_{2}^{\prime} \, Y \, \left[ \, Y_{i} \, , \, Y_{j} \, \right]_{\star} \; .
\end{eqnarray}
Thus the transformations of field strengths are
\begin{eqnarray}
X_{\mu\nu} \; \; \longmapsto \; \; X_{\mu\nu}^{\, \prime} &=& V \star X_{\mu\nu} \star V^{-1} \; ,
\nonumber \\
Y_{ij} \; \; \longmapsto \; \; Y_{ij}^{\, \, \prime} &=& V \star Y_{ij} \star V^{-1} \; .
\end{eqnarray}
Using the construction of the section $V$, we have the Lagrangian
\begin{eqnarray}
{\cal L}_{gauge}&=&-\frac{1}{2}\, \mbox{tr}\left(F_{\mu\nu} \, \star \, F^{\mu\nu}\right)
-\frac{1}{2}\, \mbox{tr}\left(G_{ij} \, \star \, G_{ij}\right)
- \mbox{tr}\left(F_{ij} \, \star \, G_{ij}\right)
\nonumber \\
&&
-\frac{1}{4} \, X_{\mu\nu} \, \star \, X^{\mu\nu}
-\frac{1}{4} \, Y_{ij} \, \star \, Y_{ij}
- \frac{1}{2} \, X_{ij} \, \star \, Y_{ij} \; .
\end{eqnarray}

The symmetry proposed here is the simplest structure of group for the strong interaction unified to
electromagnetic interaction in the scenario DFR NC in $D=4+3$. We will introduce two Higgs sectors to
reduce the symmetry, and so, we identify the final symmetry of unification $SU_{c}^{\star}(3) \times U_{em}^{\star}(1)$.


\section{The sector of Higgs for the model $U^{\star}(3) \times U^{\star}(1) \times U^{\star}(3) \times U^{\star}(1)$.}

In the previous section we have constructed a model for strong interaction based on the Yang-Mills symmetry DFR NC.
The structure of group is composite by $U^{\star}(1)_{A^{\mu \, 0}} \times SU^{\star}(3)_{A^{\mu \, a}} \times  U^{\star}(1)_{X^{\mu}} \times U^{\star}(1)_{B_{i}^{\, 0}} \times SU^{\star}(3)_{B_{i}^{\, a}} \times U^{\star}(1)_{Y_{i}}$, with two Abelian subgroups. We will use two
Higgs sectors to eliminate the residual symmetries to search the the final symmetry of model, that is,
\begin{eqnarray*}
U^{\star}(1)_{A^{\mu \, 0}} \times SU^{\star}(3)_{A^{\mu \, a}} \times  U^{\star}(1)_{X^{\mu}} \times U^{\star}(1)_{B_{i}^{\, 0}} \times SU^{\star}(3)_{B_{i}^{\, a}} \times U^{\star}(1)_{Y_{i}}
&& \, \, \stackrel{\langle \Phi \rangle}{\longmapsto} SU_{c}^{\star}(3) \times U_{em}^{\star}(1).
\end{eqnarray*}

We are going to introduce a second Higgs sector $\Phi$ in order to break the symmetry $SU(3)_{B_{i}^{\, a}}$.
We write the Lagrangian of the second Higgs-$\Phi$ as the sum of the scalar sector and the Yukawa Lagrangian, that is,
\begin{equation}\label{LHiggs1}
{\cal L}_{Higgs}=\left(D_{\mu}\Phi\right)^{\dagger} \star D^{\mu} \Phi
+ \, \left(D_{\theta i} \Phi \right)^{\dagger} \star D_{\theta i}\Phi
-\mu^2 \left(\Phi^{\dagger} \star \Phi\right)
-g_{h} \, \left(\Phi^{\dagger} \star \Phi\right)^{2} \; .
\end{equation}
where $\mu$ and $g_{h}$ are real parameters. The covariant derivative $D_{\mu}$ acts on $\Phi$
coupling it to the NC Abelian gauge fields
\begin{eqnarray}
D_{\mu} \Phi &=& \partial_{\mu} \, \Phi + i \, g_{1} \, A_{\mu}^{0} \star \Phi-i \, g_{1}^{\prime} \, X_{\Phi} \, \Phi \star X_{\mu} \; ,
\end{eqnarray}
while that in the sector of antisymmetrical fields, we have
\begin{eqnarray}
D_{\theta i} \Phi &=& \lambda \, \partial_{\theta i} \, \Phi + i \, g_{2} \, B_{i}^{0} \star \Phi
+i \, g_{2} \, B_{i}^{\, a} \, \frac{\lambda^{a}}{2} \star \Phi-i \, g_{2}^{\prime} \, Y_{\Phi} \, \Phi \star Y_{j}  \; .
\end{eqnarray}
Thus the field $\Phi$ is a complex scalar triplet that has the gauge
transformation under mixed Abelian groups given by
\begin{eqnarray}\label{transfPhi1}
\Phi=
\left(
\begin{array}{c}
\phi_{1} \\
\phi_{2} \\
\phi_{3} \\
\end{array}
\right)
\longmapsto \Phi^{\prime}=V_{1} \star \Phi \star V^{-1} \; ,
\end{eqnarray}
where $V_{1}$ is the element of the subgroup $U^{\star}(1)_{A^{\mu \, 0}}$, and the transformation of $\Phi$ with
respect to $U^{\star}(3)_{B_{i}} \times U^{\star}(1)_{Y_{i}}$ is
\begin{eqnarray}\label{transfPhi1}
\Phi \; \; \longmapsto \; \; \Phi^{\prime}=U \star \Phi \star V^{-1} \; .
\end{eqnarray}
The reader can check that the covariant derivatives have the similar transformations
\begin{eqnarray}
D_{\mu} \Phi \; \longmapsto \;  \left(D_{\mu}\Phi\right)^{\prime}&=&V_{1} \star \left(D_{\mu}\Phi\right) \star V^{-1} \; ,
\nonumber \\
D_{\theta i} \Phi \; \longmapsto \;  \left(D_{\theta i}\Phi\right)^{\prime}&=& U \star \left(D_{\theta i}\Phi\right) \star V^{-1} \; .
\end{eqnarray}
The minimal value of the previous Higgs potential can be obtained by the non-trivial vacuum expected value (VEV) of
the scalar that keeps the translational invariance in the space $x+\theta$, where we can choose it as the
triplet of the constant VEV
\begin{eqnarray}
\langle \Phi \rangle_{0}=
\left(
\begin{array}{c}
0 \\
0 \\
\frac{v}{\sqrt{2}} \\
\end{array}
\right) \; ,
\end{eqnarray}
where $v$ is the non-trivial vacuum expectation value (VEV) of the Higgs field $\Phi$, defined by $v:=\sqrt{-\mu^{2} / g_{h}}$, when $\mu^{2}<0$. We choose the parametrization of the $\Phi$-complex field as
\begin{eqnarray}\label{PhiGaugeunitary}
\Phi(x,\theta)= \frac{\left(v+H\right)}{\sqrt{2}} \star e^{i \, \frac{\lambda^{a}}{2}\frac{\chi^{a}}{v}}
\left(
\begin{array}{c}
0 \\
0 \\
1 \\
\end{array}
\right) \; ,
\end{eqnarray}
where $\chi^{a}$ and $H$ are real functions of $(x,\theta)$.
%
%
This VEV defines a scale concerning the breaking of the residual symmetry in which the NC Abelian
gauge field acquires a mass term. Since we have interested in the mass spectrum for the NC bosons,
we choose the unitary gauge condition in which $\chi^{a}=0$ by simplicity. So the free part of the Lagrangian is given by
\begin{eqnarray}\label{LHiggs1Masses}
{\cal L}_{Higgs-0}&=&\frac{1}{2}\,\left(\partial_{\mu}H\right)^{2}
+\frac{\lambda^{2}}{2}\left(\partial_{\theta i}H\right)^{2}
-\frac{1}{2}\left(-2\mu^{2}\right)H^{2}
+\frac{g_{2}^{2} \, v^{2}}{4} \, \left(B_{i}^{\,\,\,a}\right)^2
\nonumber \\
&& +\frac{v^{2}}{2} \left(\phantom{\frac{1}{2}} \!\!\!\! g_{1} \, A_{\mu}^{0}- g_{1}^{\prime} \, X_{\Phi} \, X_{\mu} \, \right)^{2}
\!+\frac{v^{2}}{2} \left( \phantom{\frac{1}{2}} \!\!\!\! g_{2} \, B_{i}^{0}-g_{2}^{\prime} \, Y_{\Phi} \, Y_{i}^{0} \, \right)^{2} \; ,
\end{eqnarray}
where we have used the identity of the Moyal product (\ref{ProductMoyal}). In this expression we have obtained
the Higgs $H$-field Lagrangian with propagation along the $\theta$-space, and a mass of $m_{H}=\sqrt{2 \, g_{h} \, v^2}$.

We observe the emergence of a mass term for the eight vector fields $B_{i}^{\, a}$ identified by
\begin{eqnarray}
m_{B_{i}^{\,\,a}}=\frac{ g_{2} \, v}{\sqrt{2}} \; .
\end{eqnarray}
In this stage, we meet all propagation terms after the gauge sector acquires VEV
\begin{eqnarray}
{\cal L}_{gauge-0}&=&-\frac{1}{4}\, \left(\partial_{\mu}A_{\nu}^{\, 0}-\partial_{\nu}A_{\mu}^{\, 0} \right)^{2}
-\frac{1}{4}\, \left(\partial_{\mu}X_{\nu}-\partial_{\nu}X_{\mu} \right)^{2}
\nonumber \\
&&
- \frac{1}{2} \, \left(\partial_{\mu}A_{\nu}^{\, 0}-\partial_{\nu}A_{\mu}^{\, 0} \right)
\, \left(\partial^{\mu}X^{\nu}-\partial^{\nu}X^{\mu} \right)
\nonumber \\
&&
+\frac{v^{2}}{2} \left(\phantom{\frac{1}{2}} \!\!\!\! g_{1} \, A_{\mu}^{0}- g_{1}^{\prime} \, X_{\Phi} \, X_{\mu} \, \right)^{2}
+\frac{v^{2}}{2} \left( \phantom{\frac{1}{2}} \!\!\!\! g_{2} \, B_{i}^{0}-g_{2}^{\prime} \, Y_{\Phi} \, Y_{i}^{0} \, \right)^{2}
\end{eqnarray}

The others mixed terms in (\ref{LHiggs1Masses}) motivate us to introduce the orthogonal transformations
\begin{eqnarray}\label{transfA0CGY}
A_{\mu}^{ \, 0} &=& \cos\alpha \, G_{\mu}+ \sin\alpha \, A_{\mu}
\nonumber \\
X_{\mu} &=& -\sin\alpha \, G_{\mu}+\cos\alpha \, A_{\mu} \; ,
\end{eqnarray}
and
\begin{eqnarray}\label{transfB0X}
B_{i}^{\, 0} &=& \cos\beta \, C_{i}+ \sin\beta \, B_{i}
\nonumber \\
Y_{i} &=& -\sin\beta \, C_{i}+\cos\beta \, B_{i} \; ,
\end{eqnarray}
where $\alpha$ and $\beta$ are the mixing angles, and it satisfy the conditions
\begin{eqnarray}\label{tanalphabeta}
\tan\alpha=\frac{g_{1}^{\prime}}{g_{1}}
\hspace{0.5cm} \mbox{and} \hspace{0.5cm}
\tan\beta=\frac{g_{2}^{\prime}}{g_{2}} \; .
\end{eqnarray}
Here we have used for convenience that $X_{\Phi}=Y_{\Phi}=1$.
Thus the Lagrangian (\ref{LHiggs1Masses}) can be written as
\begin{equation}\label{LHiggs1Masses2}
{\cal L}_{Higgs-0} = \frac{1}{2} \, \left(\partial_{\mu}H\right)^{2}
+\frac{\lambda^{2}}{2}\left(\partial_{\theta i}H\right)^{2}
-\frac{1}{2} \, m_{H}^{2} \, H^{2}
+ \, \frac{1}{2} \, m_{B_{i}^{\,a}}^{2} \left(B_{i}^{\,\, a}\right)^{2}
+\frac{1}{2} \, m_{G}^{2} \, G_{\mu}^{\, 2}
+\frac{1}{2} \, m_{C}^{2} \, C_{i}^{\, 2} \; ,
\end{equation}
where the mass of $G^{\mu}$ is given by the expression
\begin{eqnarray}\label{massGmu}
m_{G}=v \, \sqrt{g_{1}^{\, 2}+g_{1}^{ \, \prime \, 2}}=\frac{g_{1} \, v}{\cos\alpha} \; ,
\end{eqnarray}
while the mass of $C_{i}$ is given by
\begin{eqnarray}
m_{C}=v \, \sqrt{g_{2}^{\, 2}+ g_{2}^{ \, \prime \, 2}}=\frac{g_{2} \, v}{\cos\beta}  \; .
\end{eqnarray}

Finally, we can meet all the terms of the gauge sector after the SSB :
\begin{eqnarray}\label{FreeLGY}
{\cal L}_{gauge-0}&=&-\frac{1}{4} \, \left(\partial_{\mu}A_{\nu}^{\, a}-\partial_{\nu}A_{\mu}^{\, a}\right)^{2}
-\frac{1}{4} \, \left(\partial_{\mu}A_{\nu}-\partial_{\nu}A_{\mu}\right)^{2}
\nonumber \\
&&
-\frac{\lambda^{2}}{4} \, \left(\partial_{\theta i}B_{j}-\partial_{\theta j}B_{i}\right)^{2}
\nonumber \\
&&
-\frac{\lambda^{2}}{4} \, \left(\partial_{\theta i}B_{j}^{\,\, a}-\partial_{\theta j}B_{i}^{\,\, a} \right)^{2}
+ \, \frac{1}{2} \, m_{B_{i}^{\, a}}^{2} \left(B_{i}^{\,\, a}\right)^{2}
\nonumber \\
&&
-\frac{1}{4} \, \left(\partial_{\mu}G_{\nu}-\partial_{\nu}G_{\mu} \right)^{2}+\frac{1}{2} \, m_{G}^{\, 2} \, G_{\mu}^{\, 2}
\nonumber \\
&&
-\frac{\lambda^{2}}{4} \, \left(\partial_{\theta i}C_{j}-\partial_{\theta j}C_{i}\right)^{2}
+\frac{1}{2} \, m_{C}^{\, 2} \, C_{i}^{\,2}
\nonumber \\
&&+ \, \mbox{mixing terms between gauge boson fields} \; ,
\end{eqnarray}
where in this expression only the gauge fields $A^{\mu \, a}$, $B_{i}$ and $A^{\mu}$ are massless.
Here we have eight NC massless gluons $A^{\mu \, a} \, \left( \, a=1 \, , \, 2 \, , \, \ldots \, , \, 8 \, \right)$,
the massless NC photon $A^{\mu}$ and it correspondent $B^{i}$ of the extra symmetry $U^{\star}(1)_{B^{i}}$ that remains
in the model after the SSB. After this SSB, we will obtain a remaining symmetry such as
\begin{eqnarray*}
U^{\star}(1)_{A^{\mu 0}} \times SU^{\star}(3)_{A^{\mu a}} \times U^{\star}(1)_{X^{\mu}} \times  U^{\star}(1)_{B_{i}^{\,0}} \times SU^{\star}(3)_{B_{i}^{\, a}} \times U^{\star}(1)_{Y_{i}} \stackrel{\langle \Phi \rangle}{\longmapsto}
\nonumber \\
SU^{\star}(3)_{A^{\mu a}} \times U^{\star}(1)_{A^{\mu}}
\times U^{\star}(1)_{B_{i}}= SU_{c}^{\star}(3) \times U^{\star}(1)_{em} \; .
\end{eqnarray*}

In order to obtain a numerical value for the masses of $G^{\mu} , B_{i}^{\, \, a}$ and $C_{i}$ in terms of the electromagnetic interaction
and the mixing angles, we need to analyze the interactions between gauge bosons, quarks and leptons, and self-interactions of the gauge bosons.
This issue will be discussed in the next section.

\section{Interactions and numerical masses of the new bosons}

The interactions between leptons-quarks and the gauge vector bosons are given by
\begin{eqnarray}\label{LintGY}
{\cal L}_{quarks/leptons-gauge}^{\, int}&=& \bar{\ell} \star \left( \, g_{1}^{\prime} \, X_{\ell} \, \, \slash{\!\!\!\!X} \, \right) \star \ell
+ \bar{\ell} \star \gamma^{\mu} \left[ \!\!\!\!\! \phantom{\frac{1}{2}}  \, \ell \, , \, g_{1}^{\prime} \, \, X_{\mu} \, \right]_{\star}
\nonumber \\
&& -\bar{\Psi} \star \left( \!\!\!\! \phantom{\frac{1}{2}} g_{1} \, \slash{\!\!\!\!A}^{0} -g_{1}^{\prime} \, X_{\Psi} \, \slash{\!\!\!\!X}  \right)
\star \Psi
+ \bar{\Psi} \star \gamma^{\mu} \left[ \!\!\!\! \phantom{\frac{1}{2}} \Psi \, , \, g_{1}^{\prime} \, X_{\Psi} \, \, X_{\mu} \, \right]_{\star} ,
\end{eqnarray}
where $\Psi$ sets any quark of the model, that is, $\Psi=\left(q_{u} \, , \, q_{d} \, \right)$. Using (\ref{transfA0CGY}),
the previous expression can be written in terms of the fields $G^{\mu}$ and $A^{\mu}$ to show us the emergence of the electromagnetic interaction
\begin{eqnarray}\label{LintGY}
{\cal L}_{quarks/leptons-gauge}^{\, int}&=& \left( \, g_{1}^{\prime} \, \cos\alpha \, X_{\ell} \, \right) \bar{\ell} \star \, \slash{\!\!\!\!A}  \star \ell
+ \left( \, g_{1}^{\prime} \, \cos\alpha \, X_{\ell} \, \right) \bar{\ell} \star \gamma^{\mu} \left[ \!\!\!\!\! \phantom{\frac{1}{2}}  \, \ell \, ,
\, \, A_{\mu} \, \right]_{\star}
\nonumber \\
&&
-\left( \, g_{1}^{\prime} \, \sin\alpha \, X_{\ell} \, \right) \bar{\ell} \star \, \slash{\!\!\!\!G}  \star \ell
-\left( \, g_{1}^{\prime} \, \sin\alpha \, X_{\ell} \, \right) \bar{\ell} \star \gamma^{\mu} \left[ \!\!\!\!\! \phantom{\frac{1}{2}}  \, \ell \, , \, \, G_{\mu} \, \right]_{\star}
\nonumber \\
&&
- \left( \, g_{1} \, \sin\alpha - X_{\Psi} \, g_{1}^{\prime} \cos\alpha \, \right) \bar{\Psi} \star \, \slash{\!\!\!\!A}  \star \Psi
\nonumber \\
&&
+ g_{1}^{\prime} \cos\alpha  \, X_{\Psi} \, \bar{\Psi} \star \gamma^{\mu} \left[ \!\!\! \phantom{\frac{1}{2}} \Psi \, , \, \, A_{\mu} \, \right]_{\star}
\nonumber \\
&&
- \left( \, g_{1} \, \cos\alpha + X_{\Psi} \, g_{1}^{\prime} \sin\alpha \, \right) \bar{\Psi} \star \, \slash{\!\!\!\!G}  \star \Psi
\nonumber \\
&&
- g_{1}^{\prime} \, \sin\alpha \, X_{\Psi} \, \bar{\Psi} \star \gamma^{\mu} \left[ \!\!\! \phantom{\frac{1}{2}} \Psi \, , \, \, G_{\mu} \, \right]_{\star} \; .
\end{eqnarray}
The coupling constant is identified as the electric charge in the interaction lepton-photon by means the parametrization
\begin{eqnarray}
e=g_{1} \, \sin\alpha = g_{1}^{\prime} \, \cos\alpha \; ,
\end{eqnarray}
thus the generator $X_{\ell}$ is the leptons electric charge, $X_{\ell}=Q_{em}=-1$.
The electric charge of quarks is so by the relation
\begin{eqnarray}
Q_{quarks}=1-X_{quarks} \; ,
\end{eqnarray}
in which we known the charges $Q_{u}=2/3$ and $Q_{d}=-1/3$ for up and down quarks, respectively. Thereby, we can easily determinate the primordial generators as $X_{u}=+1/3$ and $X_{d}=+4/3$.

Thereby the interactions leptons/quarks with abelian gauge bosons are given by
\begin{eqnarray}\label{LintGY}
{\cal L}_{leptons-gauge}^{\, int}&=& - \, e \, \bar{\ell} \star \, \slash{\!\!\!\!A} \star \ell
- \, e \, \bar{\ell} \star \gamma^{\mu} \left[ \!\!\!\!\! \phantom{\frac{1}{2}}  \, \ell \, , \, A_{\mu} \, \right]_{\star}
\nonumber \\
&&
+ \, e \, \tan\alpha \, \bar{\ell} \star \, \slash{\!\!\!\!G} \star \ell
+ \, e \, \tan\alpha \, \bar{\ell} \star \gamma^{\mu} \left[ \!\!\!\!\! \phantom{\frac{1}{2}}  \, \ell \, , \, G_{\mu} \, \right]_{\star} \; ,
\end{eqnarray}
and
\begin{eqnarray}
{\cal L}_{quarks-gauge}^{\, int}&=& - \, \frac{2e}{3} \, \bar{q}_{u} \star \, \slash{\!\!\!\!A} \star q_{u}
+ \, \frac{e}{3} \, \bar{q}_{u} \star \gamma^{\mu}\left[ \!\!\! \phantom{\frac{1}{2}} q_{u} \, , \, A_{\mu} \right]_{\star}
\nonumber \\
&&
- \, e \, \cot\alpha \left( 1  + \frac{1}{3} \, \tan^{2}\alpha \right) \bar{q}_{u} \star \, \slash{\!\!\!\!G} \star q_{u}
\nonumber \\
&&
- \, \frac{e}{3} \tan\alpha \, \, \bar{q}_{u} \star \gamma^{\mu}\left[ \!\!\! \phantom{\frac{1}{2}} q_{u} \, , \, G_{\mu} \right]_{\star}
\nonumber \\
&&
+ \, \frac{e}{3} \, \bar{q}_{d} \star \, \slash{\!\!\!\!A} \star q_{d}
+ \, \frac{4e}{3} \, \bar{q}_{d} \star \gamma^{\mu}\left[ \!\!\! \phantom{\frac{1}{2}} q_{d} \, , \, A_{\mu} \right]_{\star}
\nonumber \\
&&
- \, e \, \cot\alpha \left( \, 1  + \frac{4}{3} \, \tan^{2}\alpha \, \right) \bar{q}_{d} \star \, \slash{\!\!\!\!G} \star q_{d}
\nonumber \\
&&
- \, \frac{4e}{3} \, \tan\alpha \, \bar{q}_{d} \star \gamma^{\mu}\left[ \!\!\! \phantom{\frac{1}{2}} q_{d} \, , \, G_{\mu} \right]_{\star} \; .
\end{eqnarray}
The interaction of the non-abelian sector $SU^{\star}(3)$ has analogous structure of the commutative case $SU(3)_{c}$
\begin{eqnarray}
{\cal L}_{SU^{\star}(3)}^{int}&=&-g_{1} \, \bar{\Psi} \star \gamma^{\mu} A_{\mu}^{\, a} \, \frac{\lambda^{a}}{2} \star \Psi
-g_{1} \, f^{abc} \, \partial_{\mu}A_{\nu}^{\, a} \left[ \, A^{\mu \, b} \, , \, A^{\nu \, c} \, \right]_{\star}
\nonumber \\
&&
-\frac{g_{1}^{\, 2}}{2} \, f^{abc} \, f^{ade} \, \left[ \, A_{\mu}^{\, b} \, , \, A_{\nu}^{\, c} \, \right]_{\star} \left[ \, A^{\mu \, d} \, , \, A^{\nu \, e} \, \right]_{\star} \; ,
\end{eqnarray}
that in the commutative limit, it reduces immediately to the case of QCD. Therefore, $g_{1}$ is identified as the constant coupling
of the sector of QCD in the model $SU^{\star}(3)\times U^{\star}(1)_{em}$. It must gain some contribution of the NC scale in the programme
of renormalization of the model. So we use the experimental value of the fine structure constant \cite{Boito2015}
\begin{eqnarray}
\alpha_{s}=\frac{g_{1}^{\, 2}}{4\pi}=0.303 \, \pm \, 0.009 \; ,
\end{eqnarray}
to estimate the mixing angle $\alpha$ as $\sin\alpha\simeq0.15$, and the other coupling constant is $g_{1}^{\prime} \simeq 0.30$.
Therefore, the mass of the boson $G^{\mu}$ is written in terms of scale VEV as
\begin{eqnarray}
m_{G}=\frac{e \, v}{\sin\alpha \, \cos\alpha} \simeq 2 \; v \; .
\end{eqnarray}
Thereby, the mass of the boson $G^{\mu}$ divided by two defines the VEV scale of this SSB. This relation can be used
to fix the VEV-scale since we know a estimative for $G^{\, \mu}$ - mass. To get it we use the mass of $G^{\mu}$ boson
so obtained in the NC electroweak model \cite{MJNevesEPL2016}
\begin{eqnarray}
m_{G^{\mu}}=770 \, \, \mbox{GeV} \; ,
\end{eqnarray}
so the $v$-scale is given by
\begin{eqnarray}
v=385 \, \, \mbox{GeV} \; .
\end{eqnarray}

To determinate the mixing angle $\beta$ of the extra-fields sector we must analyze the strength field tensors of $C_{i}$ and $B_{i}$.
Using the orthogonal transformation (\ref{transfB0X}), the tensors are consistent if we impose the following condition
\begin{eqnarray}
g_{2}^{\, \prime} \, \sin\beta = g_{2} \, \cos\beta \; ,
\end{eqnarray}
where we have adopted for convenience $Y=+1$. Thus the second expression in (\ref{tanalphabeta}) fixes the mixing angle $\beta$ at
$\tan \beta = 1$, that is, the physical solution is $\beta=45^{0}$, and consequently, $g_{2}=g_{2}^{\, \prime}$. The constant coupling
$g_{2}$ is obtained by analyzing the interactions photon-photon with the $B_{i}$ - field from (\ref{transfB0X}) :
%
%
\begin{figure}[!h]
\begin{center}
\newpsobject{showgrid}{psgrid}{subgriddiv=1,griddots=10,gridlabels=6pt}
\begin{pspicture}(-1,1.5)(20,3)
\psset{arrowsize=0.2 2}
\psset{unit=1}
%
%
%
\pscoil[coilarm=0,coilaspect=0,coilwidth=0.2,coilheight=1.0,linecolor=black](0,1)(3,1)
\pscoil[coilarm=0,coilaspect=0,coilwidth=0.15,coilheight=1.0,linecolor=black](1.45,1.1)(0.5,3)
\pscoil[coilarm=0,coilaspect=0,coilwidth=0.15,coilheight=1.0,linecolor=black](1.45,1.1)(2.5,3)
\put(0,2.8){\large$\gamma$}
\put(2.7,2.8){\large$\gamma$}
\put(-0.2,1.25){\large$B_{i}$}
\put(2.8,1.25){\large$B_{j}$}
\put(3.5,1){${\cal L}_{\gamma\gamma-BB}^{\, \, int}=-\frac{1}{2}\left( \, g_{1} \, g_{2} \sin^{2}\alpha \sin^{2}\beta+g'_{1} \, g'_{2} \cos^{2}\alpha \cos^{2}\beta \, \right) \, \left[A_{i} \, , \, A_{j}\right]_{\star}\left[B_{i} \, , \, B_{j}\right]_{\star}$}
%
%
\end{pspicture}
%
%
%
\end{center}
\end{figure}

\noindent
Comparing it with the interaction (\ref{Lint4AABB}), the coupling gives us the relation
\begin{eqnarray}
g_{2}=g'_{2}=\frac{2 \, e'}{\sin\alpha+\cos\alpha} \simeq 1.77 \, \, e^{\prime} \; .
\end{eqnarray}
We meet the masses of $B_{i}^{\, a}$ and $C_{i}$ - bosons in terms of the parameters $e'$ and $v$
\begin{eqnarray}
m_{B_{i}^{\,\,a}} & \simeq & 1.25 \, \, e^{\prime} \, v
\nonumber \\
m_{C} & \simeq & 2.5 \, \, e^{\prime} \, v  \; ,
\end{eqnarray}
and as consequence, we obtain the relation $m_{C}= 2 \, \, m_{B_{i}^{\,a}}$. To estimate the mass $m_{C}$, we interpret the
bosons $B_{i}^{\, a}$ as possible massive gluons whose the origin must be associated with glueballs \cite{Ochs2013}.
Thus the non-commutative model contains bosons family extended in that massive gluons would emerge in the extra $\theta$-space.
The spectrum of gluonium bound states $gg$ is in the mass range $1.5-1.7 \, \mbox{GeV}$, so if we use it as a mass range for
the NC boson $B_{i}^{\, a}$, the mass of $C_{i}$ is estimated as
\begin{eqnarray}
m_{C}=3-3.5 \, \, \mbox{GeV} \; .
\end{eqnarray}
Using the VEV result, the new coupling constant of the NC scenario is $e'=0.003$, and it correspondent fine structure constant is
\begin{eqnarray}
\alpha_{e^{\prime}}=\frac{e^{\prime\,2}}{4\pi} \simeq 7.16 \, \times \, 10^{-7} \; .
\end{eqnarray}
\section{Conclusions and perspectives}
In this work we believe that some new steps were provided in order to fathom the DFR formalism which is considered in the NC literature as a possible path to quantum gravity. Keeping all these quantum ideas in mind, we have considered gauge Abelian and non-Abelian fields using the recent DFR framework where the parameter that carries the noncommutative feature, $\theta^{\mu\nu}=\left( \, 0 \, , \, \theta^{ij} \, \right)$, in that we make
$\theta^{0i}=0$, represents independent degrees of freedom completing the DFR $D=7$ extended DFR space, where the phase-space has the momentum $K^{ij}$ associated with $\theta^{ij}$.

In this way we have started with a first quantized formalism, where $\theta^{ij}=\epsilon^{ijk} \, \theta^{k}$
and its canonical momentum $K^{ij}=\epsilon^{ijk} \, K^{k}$ are operators living in an extended Hilbert space.
This structure, which is compatible with the minimal canonical extension of the so-called DFR algebra,
is also invariant under an extended Poincar\'e group of symmetry, but keeping, among others, the usual Casimir invariant operators. After that, in a second quantized formalism scenario, we have succeed in presenting an explicit form for the extended Poincar\'e
generators and the same algebra of the first quantized description has been generated via generalized Heisenberg relations.  This is a fundamental  point because the usual Casimir operators for the Poincar\'e group are proven to be kept and to maintain the usual classification scheme for the elementary particles.

After that, in order to complete the DFR fermionic formalism given in \cite{Amorim2} we have constructed the Gamma matrices, its algebra and the DFR Dirac equation were also analyzed. These results set the stage to discuss the gauge invariance subject in DFR scenario where $\star$-covariant derivatives were used in order to construct the local DFR gauge transformations. These ones allow us to construct DFR gauge invariant Lagrangians for the DFR versions of the QED and Yang-Mills models. The DFR NC conjecture has revealed the existence of an extra gauge field, beyond the NC
vector gauge field, to maintain the gauge invariance in the $\theta$-space. This new gauge field is due necessarily to DFR noncommutativity.
It is easy to see that in the commutative limit, any influence of this gauge field in the model goes to zero.
The gauge symmetry is based on the composite group $U^{\star}(N) \times U^{\star}(N)$ , in which the extra $U^{\star}(N)$ is due to new NC
gauge field. We have seen that the NC electromagnetism came out naturally from these DFR version by the symmetry
$U_{em}^{\star}(1)=U^{\star}(1) \times U^{\star}(1)$. Since the gauge invariance is
consistent, it guarantees the renomalizability and unitarity of the model. It must be investigated in another paper.

After that, as an application of the DFR composite symmetry model, we have constructed a Strong Interaction Model which is unified to electromagnetism theory
based on the symmetry $U^{\star}(3) \times U^{\star}(1) \times U^{\star}(3) \times U^{\star}(1)$. The Higgs sector has been introduced
to reduce this symmetry to $SU_{c}^{\star}(3) \times U_{em}^{\star}(1)$, and as consequence, some bosons of the model have acquired mass
through a scheme of spontaneous symmetry breaking. With the help of some results of both Strong interaction and Electroweak model.  We have estimated these masses
and also the constant coupling extremely weak associated with the DFR NC new gauge bosons.

A study of the quantum aspects of this new field will motivate our research in the future. We have also obtained some quantum aspects of the Yang-Mills model which opens the possibility to investigate its quantization. The study of the divergences can reveal the possiblity of the non existence of the ultraviolet/infrared mixing in DFR Yang-Mills model in the way we know. It would simplify the renormalization procedure of the model. We can ask what is the influence of this NC scenario in the functions of the Renormalization group. The issue  about the asymptotic freedom deserves to be investigated with more details. It is an ongoing research.

\section{Acknowledgments}

\ni E.M.C.A. thanks CNPq (Conselho Nacional de Desenvolvimento Cient\' ifico e Tecnol\'ogico), Brazilian scientific support federal agency, for partial financial support, Grants No. 301030/2012-0 and No. 442369/2014-0 and the hospitality of Theoretical Physics Department at Federal University of Rio de Janeiro (UFRJ), where part of this work was carried out. The authors thank also professor J. A. Helay\"el Neto, for valuable discussions.


\begin{thebibliography}{30}

\bibitem{snyder47} H. S. Snyder, Phys. Rev. 71 (1947)  38.

\bibitem{yang47}  C. N. Yang, Phys. Rev.  72 (1947) 874.

\bibitem{seibergwitten99} N. Seiberg and E. Witten, JHEP 9909 (1999) 032.

\bibitem{QG4}   R. J. Szabo, Class. Quant. Grav. 23 (2006) R199.

\bibitem{QG5}   R. J. Szabo, {\it Quantum Gravity, Field Theory and Signatures of NC Spacetime}, arXiv:0906.2913.

\bibitem{Szabo03} R. Szabo,   Phys. Rep.   378 (2003) 207.

\bibitem{Chaichian}  M. Chaichian, M. M. Sheikh-Jabbari and A. Tureanu,   Phys. Rev. Lett.  86 (2001), 2716-2719.

\bibitem{DFR1} S. Doplicher, K. Fredenhagen and J. E. Roberts,  Phys. Lett. B   331 (1994) 29.

\bibitem{DFR2} S. Doplicher, K. Fredenhagen and J. E. Roberts,  Commun. Math. Phys.   172 (1995) 187.

\bibitem{Morita} H. Kase, K. Morita, Y. Okumura and E. Umezawa, {\it Prog. Theor. Phys.}
 {\bf 109} (2003) 663; K. Imai, K. Morita and Y. Okumura, {\it Prog. Theor. Phys.}
{\bf 110} (2003) 989; A. Deriglazov, Phys. Lett. B 555 (2003) 83.

\bibitem{Amorim1}    R. Amorim, {\it Phys. Rev.  Lett.} {\bf 101} (2008) 081602.

\bibitem{amo} E. M. C. Abreu, A. C. R. Mendes, W. Oliveira and A. Zagirolamim, {\it SIGMA} {\bf 6} (2010) 083.

\bibitem{EMCAbreuMJNeves2012} E. M. C. Abreu and M. J. Neves, {\it Int. J. Mod. Phys.} A {\bf 27} (2012) 1250109.

\bibitem{Amorim4}  R. Amorim,   Phys. Rev. D   78 (2008) 105003.

\bibitem{Amorim5}  R. Amorim,   J. Math. Phys.   50 (2009) 022303.

\bibitem{Amorim2}  R. Amorim,    J. Math. Phys.   50  (2009) 052103.

\bibitem{EMCAbreuMJNeves2013} E. M. C. Abreu and M. J. Neves, {\it Int. J. Mod. Phys.} A {\bf 28} (2013) 1350017.

\bibitem{aa} E. M. C. Abreu, R. Amorim and W. G. Ramirez, Phys. Rev. D 80 (2009) 105010.

\bibitem{EMCAbreuMJNeves2014} E. M. C. Abreu and M. J. Neves, {\it Nuclear Physics} B {\bf 884} (2014) 741.

\bibitem{MJNevesEPL2016} M. J. Neves and E. M. C. Abreu, {\it Europhysics Letters} {\bf 114} (2016) 21001.

\bibitem{Amorim3} R. Amorim, Phys. Rev. D 81 (2010) 105005.

\bibitem{Gamboa}  J. Gamboa, M. Loewe and J. C. Rojas,   Phys. Rev. D   64 (2001), 067901.

\bibitem{Kokado}   A. Kokado, T. Okamura and T. Saito,   Phys.  Rev. D   69 (2004), 125007.

\bibitem{Kijanka}  A. Kijanka and P. Kosinski,   Phys. Rev. D   70 (2004), 127702.

\bibitem{Calmet1}   X. Calmet,   Phys. Rev. D   71 (2005), 085012.

\bibitem{Calmet2}   X. Calmet and M. Selvaggi,   Phys. Rev. D   74 (2006), 037901.

\bibitem{Carlson}   C. E. Carlson, C.D. Carone and N. Zobin,   Phys. Rev. D 66 (2002) 075001.

\bibitem{Carone} C. D. Carone and H. J. Kwee,   Phys. Rev. D   73 (2006) 096005.

\bibitem{Conroy2003} J. M. Conroy, H. J. Kwee and V. Nazarayan, Phys. Rev. D {\bf 70}, 034017 (2004).

\bibitem{Saxell} S. Saxell, Phys. Lett. B   666 (2008) 486.
%
\bibitem{Boito2015} Diogo Boito, Maarten Golterman, Kim Maltman, James Osborne, and Santiago Peris, Phys. Rev. D 91 (2015) 034003.
%
\bibitem{Ochs2013} Wolfgang Ochs, Journal of Physics G : Nuclear and Particle Physics {\bf 40} (2013) 4.

%
\end{thebibliography}
\end{document}